\RequirePackage{ifpdf}
\ifpdf 
\documentclass[pdftex]{sigma}
\else
\documentclass{sigma}
\fi

\numberwithin{equation}{section}
\numberwithin{theorem}{section}
\numberwithin{proposition}{section}
\numberwithin{lemma}{section}
\numberwithin{corollary}{section}
\numberwithin{definition}{section}
\numberwithin{remark}{section}

\begin{document}

\allowdisplaybreaks

\renewcommand{\thefootnote}{$\star$}

\renewcommand{\PaperNumber}{056}

\FirstPageHeading

\ShortArticleName{Quantum Probability, Renormalization and Inf\/inite-Dimensional $*$-Lie Algebras}

\ArticleName{Quantum Probability, Renormalization\\ and Inf\/inite-Dimensional $\boldsymbol{*}$-Lie Algebras\footnote{This paper is a
contribution to the Special Issue on Kac--Moody Algebras and Applications. The
full collection is available at
\href{http://www.emis.de/journals/SIGMA/Kac-Moody_algebras.html}{http://www.emis.de/journals/SIGMA/Kac-Moody{\_}algebras.html}}}

\Author{Luigi ACCARDI~$^\dag$ and Andreas BOUKAS~$^\ddag$}

\AuthorNameForHeading{L.~Accardi and A.~Boukas}

\Address{$^\dag$~Centro Vito Volterra, Universit\`a di Roma ``Tor Vergata'', Roma I-00133, Italy}
\EmailD{\href{mailto:accardi@volterra.uniroma2.it}{accardi@volterra.uniroma2.it}}
\URLaddressD{\url{http://volterra.uniroma2.it}}

\Address{$^\ddag$~Department of Mathematics,  American College of Greece,\\
\hphantom{$^\ddag$}~Aghia Paraskevi, Athens 15342, Greece}
\EmailD{\href{mailto:andreasboukas@acgmail.gr}{andreasboukas@acgmail.gr}}

\ArticleDates{Received November 20, 2008, in f\/inal form May 16,
2009; Published online May 27, 2009}

\Abstract{The present paper reviews some intriguing connections which link together a~new renormalization technique, the theory of $*$-representations of inf\/inite dimensional $*$-Lie algebras, quantum probability, white noise and stochastic calculus
and the theory of classical and quantum inf\/initely divisible processes.}

\Keywords{quantum probability; quantum white noise; inf\/initely divisible process; quantum decomposition; Meixner classes; renormalization; inf\/inite dimensional Lie algebra;  central extension of a Lie algebra}

\Classification{60H40; 60G51; 81S05; 81S20; 81S25; 81T30; 81T40}

\section{Introduction}

The investigations on the stochastic limit of quantum theory in \cite{[AcLuVo02]} led to the development of quantum white noise calculus as a natural generalization of classical and quantum stochastic calculus. This was initially developed at a pragmatic level, just to the extent needed to solve the concrete physical problems which stimulated the birth of the theory \cite{[AcLuOb96],[AcLuVo95b],[AcLuVo02], [AcVo97]}.
The f\/irst systematic exposition of the theory is contained in the paper
\cite{[AcLuVo99]} and its full development in \cite{[Ayed06]}.

This new approach naturally suggested the idea to generalize stochastic calculus by extending it to higher powers of (classical and quantum) white noise. In this sense we speak of {\it nonlinear white noise calculus}.

This attempt led to unexpected connections between mathematical objects and results emer\-ged in dif\/ferent f\/ields of mathematics and at dif\/ferent times, such as white noise,
the representation theory of certain famous Lie algebras,
the renormalization problem in physics, the theory of independent
increment stationary (L\'evy) processes and in particular the Meixner classes, $\dots$.
The present paper gives an overview of the path which led to these connections.

Our emphasis will be on latest developments and open problems which are related to the renormalized powers of white noise of degree $\geq 3$ and the associated Lie algebras. We also brief\/ly review the main results in the case of powers $<3$
and, since the main results are scattered in several papers, spanning a rather long time period, we include some bibliographical references which allow the interested reader to reconstruct this development.

The content of the paper is the following.
In Section~\ref{st-prob} we state the Lie algebra renormalization problem taking as a model the Lie algebra of dif\/ferential operators with polynomial coef\/f\/icients.

In Section~\ref{section3} we recall the basic notions on $\ast$-representations of $\ast$-Lie algebras and their connections with quantum probability. Section~\ref{sec-Fk-rep} describes the standard Fock representation (i.e.\ for f\/irst order f\/ields).
Sections~\ref{curr-alg} and~\ref{Prob-impl} recall some notions on current algebras and their connections with (boson) independent increment processes.
The transition from f\/irst to second quantization (in the usual framework of Heisenberg algebras) can be considered, from the algebraic point of view, as a transition from a $\ast$-Lie algebra to its current algebra over $\mathbb R$ (or $\mathbb R^d$) and, from the probabilistic point of view, as  a transition from a class of inf\/initely divisible random variables to the associated independent increment process. These sections generalize this point of view to Lie algebras more general than the Heisenberg one.
Section~\ref{sec-quadr-pow} illustrates the role of renormalization in the quadratic case and shows how, after renormalization, the above mentioned  connection between $\ast$-Lie algebras and independent increment processes, can be preserved, leading to interesting new connections.
We also quickly describe results obtained in this direction (giving references to existing surveys for more detailed information).
Starting from Section~\ref{sec-HPWN-schm} we begin to discuss the case of higher (degree $\geq 3$) powers of white noise and we illustrate the ideas that eventually led to the identif\/ication of the
RHPWN and the $w_{\infty}$-$*$-Lie algebras (more precisely their closures) (Section~\ref{RHPWN-winf}). We illustrate the content of the no-go theorems and of the various attempts made to overcome the obstructions posed by them. The f\/inal sections outline
some connections between renormalization and central extensions and some recent results obtained in this direction.

\section[Renormalization and differential operators with polynomial coefficients]{Renormalization and dif\/ferential operators\\ with polynomial coef\/f\/icients}\label{st-prob}

 Since the standard white noises, i.e.\ the distribution derivatives of Brownian motions,
are the prototypes of free quantum f\/ields, the program to f\/ind a meaningful way
to def\/ine higher powers of white noise is related to an old open problem in physics:
the renormalization problem. This problem consists in the fact that these higher powers
are strongly singular objects and there is no unique way to attribute a meaning to them.
That is why one speaks of {\it renormalized higher powers of white noise} (RHPWN).

The renormalization problem has an old history and we refer to \cite{[BogSh80]} for a review and bibliographical indications. In the past 50 years the meaning of the term
{\it renormalization} has evolved so as to include a multitude of dif\/ferent procedures.
A common feature of all these generalizations is that, in the transition from
a discrete system to a continuous one, certain expressions become meaningless and
one tries to give a mathematical formulation of the continuous theory which {\it keeps as many properties as possible from the discrete approximation}.
One of the main dif\/f\/iculties of the problem consists in its precise formulation,
in fact one does not know a priori which properties of the discrete approximation
will be preserved in the continuous theory.

The discovery of such properties is one of the main problems of the theory.

The present paper discusses recent progresses in the specif\/ication of this problem to
a basic mathematical object: the $*$-algebra of dif\/ferential operators
with polynomial coef\/f\/icients (also called the {\it full oscillator algebra}).

More precisely, in the present paper, when speaking of the {\it renormalization problem},
we mean the following:
\begin{enumerate}\itemsep=0pt

\item[$(i)$] to construct a {\it continuous analogue} of the $*$-algebra of dif\/ferential operators
with poly\-nomial coef\/f\/icients acting on the space ${\cal C}^\infty(\Bbb R^n;\Bbb C)$
of complex valued smooth functions in $n\in\Bbb N$ real variables (here {\it continuous}
means that the space $\Bbb R^n\equiv \{ \hbox{functions} \ \{1,\dots,n\}\to \Bbb R \}$ is
replaced by the space $\{ \hbox{functions} \ \Bbb R\to \Bbb R$ \});

\item[$(ii)$] to construct a $*$-representation of this algebra as operators on a Hilbert
space ${\cal H}$ (all spaces considered in this paper will be complex separable
and all associative algebras will have an identity, unless otherwise stated);

\item[$(iii)$] the ideal goal would be to have a unitary representation, i.e.\ one
in which the skew symmetric elements of this $*$-algebra can be exponentiated,
leading to strongly continuous 1-parameter unitary groups.
\end{enumerate}

In the physical interpretation, ${\cal H}$ would be the state space of a physical system with inf\/initely many degrees of freedom (typically a f\/ield, an inf\/inite volume gas, $\dots$) and the 1-parameter unitary groups correspond to time evolutions.

The $*$-algebra of dif\/ferential operators in $n$ variables with polynomial
coef\/f\/icients can be thought of as a realization of the universal enveloping algebra
of the $n$-dimensional Heisenberg algebra and its Lie algebra structure is uniquely
determined by the Heisenberg algebra. In the continuous case the renormalization problem arises from this interplay between the structure of Lie algebra and that of associative algebra.
The developments we are going to describe were originated by the idea,
introduced in the f\/inal part of the paper \cite{[AcLuVo99]}, of {\it first renormalizing
the Lie algebra structure (i.e.\ the commutation relations),
thus obtaining a new $*$-Lie algebra,
then proving existence of nontrivial Hilbert space representations}.

In the remaining of this section we give a precise formulation
of problem $(i)$ above and explain where the dif\/f\/iculty is. In Section~\ref{Prob-impl} we explain the connection with probability, in particular white noise and other independent increment processes.

\subsection{1-dimensional case}\label{sec-1-d-cse}

The {\it position operator} acts on ${\cal C}^\infty(\Bbb R;\Bbb C)$
as multiplication by  the independent variable
\[
(qf)(x):=xf(x), \qquad x\in\Bbb R  ,\qquad f\in{\cal C}^\infty(\Bbb R;\Bbb C).
\]
The usual derivation $\partial_x$ also acts on ${\cal C}^\infty(\Bbb R;\Bbb C)$
and the two operators satisfy the commutation relation
\[
[q,\partial_x]=-1,
\]
where $1$ denotes the identity operator on ${\cal C}^\infty(\Bbb R;\Bbb C)$. Def\/ining the {\it momentum operator} by
\[
p:={\frac{1}{i}}\,\partial_x, \qquad
(pf)(x):={\frac{1}{i}} \left({\frac{df}{dx}}\right)(x)
\]
one obtains the Heisenberg commutation relations
\begin{equation*}
[q,p]=i , \qquad [q,q]=[p,p]=0,
\end{equation*}
which give a  structure of Lie algebra to the vector space
generated by the  operators $q$, $p$, $1$. This is the {\it Heisenberg algebra}
${\rm Heis}(\Bbb R)$. The associative algebra ${\cal A}(\Bbb R)$, algebraically generated by the operators $p$, $q$, $1$, is the $*$-algebra of dif\/ferential operators with polynomial coef\/f\/icients
in one real variable and coincides with the vector space
\[
\sum_{n\in\mathbb{N}} P_n(q)p^n,
\]
 where the $P_n(X)$ are polynomials of arbitrary degree in the indeterminate $X$ and almost all the $P_n(X)$ are zero.

There is a unique complex involution $*$ on this algebra such that
\begin{equation}\label{1d-inv}
q^*=q ,\qquad p^*=p ,\qquad 1^*=1.
\end{equation}
The $*$-Lie algebra structure on ${\cal A}(\Bbb R)$, induced by the commutator, is uniquely determined by the same structure on the Heisenberg algebra and gives the commutation relations
\begin{equation}\label{HOCCR}
[p^n,q^k]=\sum^{n}_{h=1}(-i)^h \binom{n}{h} k^{(h)}q^{k-h}p^{n-h},
\end{equation}
where $k^{(h)}$ is the Pochammer symbol:
\begin{equation}\label{Poch-symb}
x^{(0)}=1, \qquad x^{(y)}=x(x-1)\cdots (x-y+1), \qquad x^{(y)}=0  \qquad  \hbox{if} \quad  y>x.
\end{equation}

\subsection{Discrete case}\label{sec-discr-cse}

Let us f\/ix $N\in\Bbb N$ and replace $\Bbb R$ in Section~\ref{sec-1-d-cse} by the function space
\[
x\equiv (x_1,\dots,x_N)\in \{ \hbox{functions} \; \{1,\dots,N\} \to \Bbb R\}
=: {\cal F}_{\{1,\dots,N\}}(\Bbb R)\equiv\Bbb R^N.
\]  Thus ${\cal C}^\infty(\Bbb R;\Bbb C)$ is replaced by ${\cal C}^\infty(\Bbb R^N;\Bbb C)$. The notation ${\cal F}_{\{1,\dots,N\}}(\Bbb R)$ is more appropriate than~$\Bbb R^N$
because it emphasizes its algebra structure (for the pointwise operations), which
is required in its interpretation as a test function space (see the end of this section). For the values of a function
$x\in{\cal F}_{\{1,\dots,N\}}(\Bbb R)$ we will use indif\/ferently
the notations $x_s$ or $x(s)$.

The position and momentum operators are then
\begin{gather*}
(q_sf)(x) :=x_sf(x), \qquad s\in\{1,\dots,N\},\\
p_s := {\frac{1}{i}} {\frac{\partial}{\partial x_s}}, \qquad
(p_sf)(x)={\frac{1}{i}} \left({\frac{\partial f}{\partial x_s}}\right)(x)
, \qquad s\in\{1,\dots,N\}  , \qquad f\in{\cal C}^\infty(\Bbb R^N;\Bbb C),
\end{gather*}
which give the Heisenberg commutation relations
\begin{equation}\label{dicr-Heis}
[q_s,p_t]=i\delta_{s,t}\cdot 1, \qquad [q_s,q_t]=[p_s,p_t]=0
, \qquad s,t\in\{1,\dots,N\},
\end{equation}
where $\delta_{s,t}$ is the Kronecker delta.
The involution is def\/ined as in (\ref{1d-inv}), i.e.
\begin{equation}\label{discr-inv}
q_s^*=q_s,\qquad p_s^*=p_s,\qquad 1^*=1, \qquad s\in\{1,\dots,N\}.
\end{equation}
The vector space generated by the operators $q_s$, $p_t$, $1$ ($s,t\in\{1,\dots,N\}$)
has therefore a structure of $*$-Lie algebra, it is called the
{\it $N$-dimensional Heisenberg algebra} and is denoted
\[
{\rm Heis}(\Bbb R^N)={\rm Heis}({\cal F}_{\{1,\dots,N\}}(\Bbb R)).
\]
The associative $*$-algebra ${\cal A}(\Bbb R^N)={\cal A}({\cal F}_{\{1,\dots,N\}}(\Bbb R))$,
algebraically generated by the operators $p_s$, $q_s$, $1$ ($s,t\in\{1,\dots,N\}$),
is the $*$-algebra of dif\/ferential operators with polynomial
coef\/f\/icients in $N$ real variables and coincides with the vector space
\begin{equation}\label{norm-form-pq}
\sum_{n\in\mathbb{N}^N} P_n(q)p^n, \qquad p^0:=q^0:=1,
\end{equation}
where the $P_n(X)= P_n(X_1,\dots ,X_N)$ are polynomials of arbitrary degree
in the $N$ commuting indeterminates $X_1,\dots ,X_N$, almost all the $P_n(X)$
are zero and, for $n= (n_1,\dots ,n_N)\in\mathbb{N}^N$, by def\/inition
\[
p^n := p_1^{n_1}p_2^{n_2}\cdots p_N^{n_N}.
\]
The operation of writing the product of two such operators in the form (\ref{norm-form-pq})
can be called {\it the normally ordered form} of such an operator with respect to the
generators $p_s$, $q_s$, $1$  ($s,t\in\{1,\dots,N\}$).

Also in this case the $*$-Lie algebra structure on
${\cal A}({\cal F}_{\{1,\dots,N\}}(\Bbb R))$, induced by the commutator,
is uniquely determined by the same structure
on the Heisenberg algebra and gives the commutation relations
\begin{equation}\label{discr-HOCCR}
[p^n_s,q^k_t]=\sum^{n}_{h=1}(-i)^h
\binom{n}{h} k^{(h)}\delta^h_{s,t}q^{k-h}_tp^{n-h}_s.
\end{equation}
Notice that, with respect to formula (\ref{HOCCR}), the new ingredient is
the factor $\delta^h_{s,t}$, i.e.\ the $h$-th power of the Kronecker delta. Since
\begin{equation}\label{pow-Kron-d}
\delta^h_{s,t}=\delta_{s,t}
\end{equation}
the power is useless, but we kept it to keep track of the number of
commutators performed and to make easier the comparison with the continuous case
(see Section~\ref{sec-Bos-alg} below).
Considering ${\cal F}_{\{1,\dots,N\}}(\Bbb R)$ as a test function space and def\/ining
the smeared operators
\[
p^nq^k(f) := \sum_{t\in \{1,\dots,N\}}p^n_tq^k_t f(t)
,\qquad f\in{\cal F}_{\{1,\dots,N\}}(\Bbb R)
\]
and the scalar product
\[
 \langle f,g\rangle :=  \sum_{t\in \{1,\dots,N\}}f(t)g(t)
\]
the commutation relations (\ref{dicr-Heis}) and (\ref{discr-HOCCR}) become respectively
\begin{gather*}
[q(f),p(g)]=i \langle f,g\rangle
, \qquad [q(f),q(g)]=[p(f),p(g)]=0,
\\
[p^n(f),q^k(g)]=\sum^{n}_{h=1}(-i)^h\binom{n}{h} k^{(h)}
q^{k-h}p^{n-h}(fg).
\end{gather*}
 Notice that the algebra structure on the test function space is required only when $n+k\geq 3$.
The above construction can be extended to the case of an arbitrary discrete set $I$:
all the above formulae continue to hold with the set $\{1,\dots,N\}$ replaced by a generic
f\/inite subset of $I$ which might depend on the test function.

\subsection{Continuous case}\label{sec-Bos-alg}

In this section the discrete space $\{1,\dots,N\}$ is replaced by $\Bbb R$
and the discrete test function algebra ${\cal F}_{\{1,\dots,N\}}(\Bbb R)$
by an algebra ${\cal F}_{\Bbb R}(\Bbb R)$ of smooth functions from $\Bbb R$ into itself
(for the pointwise operations).

As an analogue of the space ${\cal C}^\infty(\Bbb R^N;\Bbb C)$ one can take
the space ${\cal C}^\infty({\cal F}_{\Bbb R}(\Bbb R);\Bbb C)$ of all smooth
 cylindrical functions on ${\cal F}_{\Bbb R}(\Bbb R)$ (i.e.\ functions
$f:{\cal F}_{\Bbb R}(\Bbb R)\to{\Bbb C}$ for which there exist $N\in\Bbb N$,
$\{s_1,\dots ,s_N\}\subset \Bbb R$, and a smooth function
$f_{s_1,\dots ,s_N}:\Bbb R^N\to \Bbb C$, such that
$f(x)=f_{s_1,\dots ,s_N}(x_{s_1},\dots$, $x_{s_N})$,
$\forall \, x\in {\cal F}_{\Bbb R}(\Bbb R)$).
This space is suf\/f\/icient for algebraic manipulations but it is too narrow to
include the simplest functionals of interest for the applications in physics or
in probability theory: this is where white noise and stochastic calculus play a role.
The position operators can be def\/ined as before, i.e.
\[
(q_sf)(x)=x_sf(x),\qquad x\in{\cal F}_{\Bbb R}(\Bbb R)
,\qquad f\in{\cal C}^\infty({\cal F}_{\Bbb R}(\Bbb R);\Bbb C).
\]
The continuous analogue of the partial derivatives ${\frac{\partial}{\partial x_s}}$,
hence of the momentum operators $p_s$, can be def\/ined by f\/ixing a subspace
${\cal F}^0_{\Bbb R}(\Bbb R)$ of
${\cal F}_{\Bbb R}(\Bbb R)$ and considering functions
$f\in{\cal C}^\infty({\cal F}_{\Bbb R}(\Bbb R);\Bbb C)$ whose
Gateaux derivative in the direction $S$
\[
D_Sf(x)=\lim_{\varepsilon\to0}{\frac{1}{\varepsilon}} (f(x+\varepsilon S)-f(x))
=\langle f'(x),S\rangle
\]
exists for any test function $S\in{\cal F}^0_{\Bbb R}(\Bbb R)$ and is a continuous
linear functional on ${\cal F}^0_{\Bbb R}(\Bbb R)$
(in some topology whose specif\/ication is not relevant for our goals).
Denoting $\langle {\cal F}^0_{\Bbb R}(\Bbb R)',{\cal F}^0_{\Bbb R}(\Bbb R)\rangle$
the duality specif\/ied by this topology, the elements of ${\cal F}^0_{\Bbb R}(\Bbb R)'$
can be interpreted as distributions on $\Bbb R$ and symbolically written in the form
\[
\langle f'(x),S\rangle = \int f'(x)(s)S(s)ds
,\qquad S\in{\cal F}^0_{\Bbb R}(\Bbb R).
\]
The distribution
\begin{equation}\label{cont-part-der}
f'(x)(s)=:{\frac{\partial f}{\partial x_s}}(x)
\end{equation}
is called the {\it Hida derivative} of $f$ at $x$ with respect to $x_s$.
Intuitively one can think of it as the Gateaux derivative along the
$\delta$-function at $s$, $\delta_s(t):= \delta(s-t)$:
\[
{\frac{\partial f}{\partial x_s}} =D_{\delta_s}f.
\]
There is a large literature on the theory of Hida distributions and we refer to
\cite{[Hida92]} for more information.
The momentum operators are then def\/ined by
\[
p_s={\frac{1}{ i}} {\frac{\partial}{\partial x_s}}={\frac{1}{ i}} D_{\delta_s }
\]
and one can prove that the following generalization of the Heisenberg
commutation relations holds:
\begin{equation}\label{cont-Heis}
[q_s,p_t]=i\delta (s-t)\cdot 1
, \qquad [q_s,q_t]=[p_s,p_t]=0
, \qquad s\in\{1,\dots,N\},
\end{equation}
where now $\delta (s-t)$ is Dirac's delta and all the identities are meant
in the usual sense of operator valued distributions, i.e.\ one f\/ixes a space
of test functions, def\/ines the smeared operators
\[
q(f) := \int_{\Bbb R} f(t)q_t dt
\]
and interprets any distribution identity as a shorthand notation for the identity
obtained by multiplying both sides by one test function for each free variable and
integrating over all variab\-les.

The involution is def\/ined as in (\ref{discr-inv}) with the only dif\/ference that now
$s\in \Bbb R$ and the vector space generated by the operator valued distributions
$q_s$, $p_t$, $1$ ($s,t\in\Bbb R$)
has therefore a structure of $*$-Lie algebra. This algebra plays a crucial role
in quantum f\/ield theory and is called the {\it current algebra of ${\rm Heis}(\Bbb R)$
over $\Bbb R$} (see Section~\ref{curr-alg}) or simply the Boson algebra over $\Bbb R$.
In the following, when no confusion can arise, we will use the term {\it Boson algebra}
also for the discrete Heisenberg algebra (\ref{dicr-Heis}).

\begin{remark}
One can combine the discrete and continuous case by considering current algebras of
${\rm Heis}(\mathbb{R})$ over
\[
\{1,\dots ,N\}\times\{1,\dots ,d\}\times \mathbb{R}\equiv \{1,\dots ,N\}\times\mathbb{R}^d,
\]
 so that the commutation relations (\ref{cont-Heis}) become
($j,k\in\{1,\dots,N\} ; x,y\in\mathbb{R}^d$)
\begin{equation*}
[q_j(x),p_k(y)]= \delta_{j,k}i\delta (x-y)\cdot 1
,\qquad  [q_j(x),q_k(y)]=[p_j(x),p_k(y)]=0.
\end{equation*}
This corresponds to considering $N$-dimensional vector f\/ields on $\mathbb{R}^d$
rather than scalar f\/ields on $\mathbb{R}$. The value of the dimension $d$ plays a
crucial role in many problems, but not in those discussed in the present paper.
Therefore we restrict our discussion to the case $d=1$. It is however important
to keep in mind that all the constructions and statements in the present paper
remain true, with minor modif\/ications, when $\mathbb{R}$ is replaced by $\mathbb{R}^d$.
\end{remark}

Up to now the discussion of the continuous case has been exactly parallel to the discrete case. Moreover some unitary representations of the continuous analogue of the Heisenberg algebra are known (in fact very few: essentially only Gaussian -- quasi-free
in the terminology used in physics, see Section~\ref{sec-quadr-pow} below).

However the attempt to build the continuous analogue of the algebra of higher
order dif\/fe\-ren\-tial operators with polynomial coef\/f\/icients leads to some principle dif\/f\/iculties. For example the naive way to def\/ine the {\it second Hida derivative}
of $f$ at $x$ with respect to $x_s$ (i.e.\ ${\frac{\partial^2 f}{\partial x_s^2}}(x)$)
would be to dif\/ferentiate the ``function'' $x\mapsto f'(x)(s)$ for f\/ixed $s\in\Bbb R$,
but even in the simplest examples, one can see that the identity
(\ref{cont-part-der}) def\/ines a distribution so that this map is meaningless.

One might try to forget the concrete realization in terms of multiplication
operators and derivatives and to generalize to the continuous case the
$*$-Lie algebra structure of ${\cal A}({\cal F}_{\{1,\dots,N\}}(\Bbb R)).\!$
This can be done for some subalgebras. For example, if the subspace
${\cal F}^0_{\Bbb R}(\Bbb R)$ is an algebra for the pointwise operations,
then one can extend the $*$-Lie algebra structure of the Boson algebra over $\Bbb R$
to f\/irst order dif\/ferential operators (vector f\/ields) by introducing functions
of the position operator, which are well def\/ined
for any test function $v\in {\cal F}^0_{\Bbb R}(\Bbb R)$ by
\[
(v(q_s)f)(x) := v(x_s)f(x),\qquad x\in{\cal F}_{\Bbb R}(\Bbb R)
, \qquad f:{\cal F}_{\Bbb R}(\Bbb R)\to \Bbb C
\]
and using the commutation relation
\begin{equation}\label{comm-vq-p}
[v(q_s),p_t]=i\delta (s-t)v'(q_s)\cdot 1
\end{equation}
which leads to
\[
[u(q_{s_1})p_{t_1},v(q_{s_2})p_{t_2}]=
iu(q_{s_1})v'(q_{s_2})\delta(t_1-s_2)p_{t_2}-iv(q_{s_2})u'(q_{s_1})\delta(s_1-t_2)p_{t_1}.
\]
In terms of test functions and with the notations:
\[
u(q;a):=\int_{\Bbb R} a_su(q_s)ds, \qquad p(b) := \int b(s)p_sds
,  \qquad a,b\in{\cal F}^0_{\Bbb R}(\Bbb R)
\]
the above commutator becomes, with $a,b,c,d\in{\cal F}^0_{\Bbb R}(\Bbb R)$:
\begin{equation}\label{Lie-alg-vect-f}
[u(q;a)p(b),v(q;c)p(d)]
=iv'(q;bc)u(q;a)p(d)-iu'(q;ad)v(q;c)p(b).
\end{equation}
Another interesting class of subalgebras is obtained by considering
the vector space generated by arbitrary (smooth) functions of $q$ and f\/irst
order polynomials in $p$. The test function form of~(\ref{comm-vq-p}) is then
\begin{equation}\label{Lie-alg-pol-q-p}
[u(q;a),p(b)] = iu'(q;ab)
\end{equation}
 which shows that, for any $n\in\mathbb{N}$, the vector space generated by the family
$\{u(q;a),p(b)\}$ where $u$ is a complex polynomial of degree $\leq n$ and $a$, $b$ are
arbitrary test functions, is a nilpotent $*$-Lie algebra. We will see in Section~\ref{ce} that the simplest nonlinear case (i.e.\ $n=2$) corresponds to the current algebra
on the unique nontrivial central extension of the one dimensional Heisenberg algebra.

The right hand sides of (\ref{Lie-alg-vect-f}) and (\ref{Lie-alg-pol-q-p}) are well
def\/ined so at least we can speak of the $*$-Lie algebra of vector f\/ields in
continuously many variables, even if we do not know if some~$*$- or unitary
representations of this algebra can be built.
In the case of the algebra corresponding to~(\ref{Lie-alg-vect-f}), one can build
unitary representations but the interpretation of these representations is still
under investigation.

The situation is dif\/ferent with the continuous analogue of the higher order
commutation relations (\ref{discr-HOCCR}) (i.e.\ when $p_t$ enters with a power $\geq 2$).
Here some dif\/f\/iculties arise even at the Lie algebra level.

In fact the continuous analogue of these relations leads to
\begin{equation}\label{cont-HOCCR}
[p^n_s,q^k_t]=\sum^{n}_{h=1}(-i)^h
\binom{n}{h} k^{(h)}\delta (s-t)^hq^{k-h}_tp^{n-h}_s,
\end{equation}
which is meaningless because it involves powers of the delta function.

Any rule to give a meaning to these powers in such a way that the brackets,
def\/ined by the right hand side of (\ref{cont-HOCCR}) induce a $*$-Lie algebra
structure, will be called a {\it renormalization rule}.

There are many inequivalent ways to achieve this goal. Any Lie algebra obtained
with this procedure will be called a {\it renormalized higher power of white noise
(RHPWN) $*$-Lie algebra}.

One might argue that, since in the discrete case the identity (\ref{pow-Kron-d}) holds,
a natural continuous analogue of the commutation relations (\ref{discr-HOCCR})
should be
\[
[p^n_s,q^k_t]=\sum^{n}_{h=1}(-i)^h
\binom{n}{h} k^{(h)}\delta (s-t)q^{k-h}_tp^{n-h}_s.
\]
We will see in Section~\ref{sec-quadr-pow} that this naive approach corresponds,
up to a multiplicative constant, to Ivanov's renormalization or to consider the current
algebra, over $\mathbb{R}$, of the universal enveloping algebra of ${\rm Heis}(\Bbb R)$.

One of the new features of the renormalization problem, brought to light by the present investigation, is that some subtle algebraic obstructions (no-go theorems) hamper this
idea at least as far as the Fock representation is concerned
(a discussion of this delicate point is in Section~\ref{nogo-thm-Iv-nrm}).

A f\/irst nontrivial positive result in this programme is that,
by separating f\/irst and second po\-wers, one can overcome these obstructions in the case
$n=2$ and the results are quite encouraging (see Section~\ref{sec-quadr-pow}).

However such a separation becomes impossible for $n\geq 3$. In fact, in Section~\ref{pers-no-go-thms} we will provide strong evidence in support of the thesis
that any attempt to force this separation at the level of a Fock type representation,
brings back either to the f\/irst or to the second order case.

\section[$*$-representations of $*$-Lie algebras: connections with quantum probability]{$\boldsymbol{*}$-representations of $\boldsymbol{*}$-Lie algebras:\\ connections with quantum probability}\label{section3}

Suppose that one f\/ixes a renormalization and def\/ines a RHPWN $*$-Lie algebra
in the sense specif\/ied above. Then, according to the programme formulated in Section~\ref{st-prob}, the next step is to build  $*$-representations of it.
Since dif\/ferent Lie algebra structures will arise from dif\/ferent renormalization procedures,
we recall in this section some notions concerning $*$-representations of general
$*$-Lie algebras and their connections with quantum probability.

\begin{definition}\label{*-rep-Lie-a}
A $*$-representation of a $*$-Lie algebra ${\cal G}$ is a triple
\[
\{ {\cal H},{\cal H}_0,\pi \},
\]
where ${\cal H}$ is a Hilbert space,
${\cal H}_0$ is a dense sub-Hilbert space of ${\cal H}$,
 $\pi$ is a representation of ${\cal G}$ into the linear operators from ${\cal H}_0$ into itself (this implies in particular that the brackets are well def\/ined on ${\cal H}_0$),
and the elements of $\pi({\cal G})$ are adjointable linear operators from ${\cal H}_0$
into itself satisfying
\begin{equation*}
\pi(l)^* = \pi(l^* ), \qquad \forall \, l\in {\cal G}.
\end{equation*}

If moreover the (one-mode) $\pi$-f\/ield operators
\begin{equation*}
F_X(z):= {\frac{1}{i}}(z\pi(X)^* - \overline z \pi(X))
\end{equation*}
are essentially self-adjoint, the $*$-representation
$\pi $ is called unitary.

A vector $\Phi \in{\cal H}$ is called cyclic for the representation $\pi$ if:
\begin{enumerate}\itemsep=0pt
\item[$(i)$] $\forall \, n\in \mathbb{N}$ the vector
\begin{equation}\label{gen-nbr-vects}
\pi(l)^n\Phi \in{\cal H}
\end{equation}
is well def\/ined (this is always the case if $\Phi \in{\cal H}_0 $);

\item[$(ii)$] denoting ${\cal H}_0(\Phi)$ the algebraic linear span of the vectors
(\ref{gen-nbr-vects}) the triple $\{ {\cal H},{\cal H}_0(\Phi),\pi \}$ is a~$*$-representation of ${\cal G}$.
\end{enumerate}
\end{definition}

\begin{remark} If $\{ {\cal H},{\cal H}_0(\Phi),\pi \}$ is a $*$-representation of ${\cal G}$
with cyclic vector $\Phi$, one can always assume that ${\cal H}_0= {\cal H}_0(\Phi)$.
In this case we omit ${\cal H}_0$ from the notations and speak only of
{\it the cyclic $*$-representation} $\{ {\cal H},\Phi,\pi \}$.

Any cyclic $*$-representation of ${\cal G}$ induces a state $\varphi$
(positive, normalized linear functional)
on the universal enveloping $*$-algebra $U({\cal G})$ of ${\cal G}$, namely
\[
\varphi (a) := \langle\Phi,\pi_U(a)\Phi\rangle, \qquad a\in U({\cal G}),
\]
where $\pi_U$ is the $*$-representation of $U({\cal G})$ induced by $\pi$.
Conversely, given a state $\varphi$ on $U({\cal G})$, the GNS construction gives a cyclic
$*$-representation of $U({\cal G})$ hence of~${\cal G}$.
Thus the problem to construct (non trivial) cyclic $*$-representations of ${\cal G}$
(we will only be interested in this type of representations) is equivalent to that
of constructing (nontrivial) states on $U({\cal G})$ hence of ${\cal G}$.
This creates a deep connection with quantum probability. To clarify these connections
let us recall (without comments, see~\cite{[Ac00c]} for more informations)
the following three basic notions of quantum probability:

\begin{definition}\label{APS-SP}\qquad
\begin{enumerate}\itemsep=0pt
\item[$(i)$] An {\it algebraic probability space} is a pair $({\cal A}, \varphi )$ where ${\cal A}$
is an (associative) $*$-algebra and $\varphi$ a state on ${\cal A}$.

\item[$(ii)$] An {\it operator process} in the algebraic probability space $({\cal A}, \varphi )$
is a self-adjoint family $G$ of algebraic generators of ${\cal A}$ (typically a set of generators of ${\cal A}$).

\item[$(iii)$] For any $n\in\mathbb{N}$ and any map $g:k\in\{1, \dots, n\}\to g_k\in G$
the complex number
\[
\varphi\left( g_{1}g_{2} \cdots g_{n}\right)
\]
is called a mixed moment of the process $G$ of order $n$.
\end{enumerate}
\end{definition}

In the above terminology one can say that  constructing a $*$-representation of
a $*$-Lie algebra~${\cal G}$ is equivalent to constructing an algebraic probability space
$\{U({\cal G}),\varphi\}$ based on the universal enveloping $*$-algebra
$U({\cal G})$ of ${\cal G}$ or equivalently, by the Poincar\'e--Birkhof\/f--Witt theorem,
an {\it operator process} in $\{U({\cal G}),\varphi\}$ given by any self-adjoint
family $G$ of algebraic generators of ${\cal G}$.

In the following section we show that when ${\cal G}$ is the Boson algebra and
$\varphi$ the Fock state, the resulting algebraic probability space
is that of the standard quantum white noise and its restriction to appropriate
maximal Abelian (Cartan) subalgebras, gives the standard classical white noise.
\end{remark}

\section{The Fock representation of the Boson algebra and white noise}\label{sec-Fk-rep}

In the present section we discuss $*$-representations of the Boson algebra
introduced in Section~\ref{sec-Bos-alg}. All the explicitly known representations
of this algebra can be constructed from a single one: the Fock representation.

To def\/ine this representation it is convenient to replace the generators $q_s$, $p_t$,
$1$  of the Boson algebra by a new set of generators $b_s^ +$ (creator), $b_t$
(annihilator), $1$ (central element, often omitted from notations) def\/ined by
\begin{equation}\label{df-aa+}
b_t^ +  =  {\frac{1 }{ \sqrt 2 }} (q_t  -  ip_t), \qquad
b_t = {\frac{1}{ \sqrt 2} }  (q_t    +  ip_t ).
\end{equation}
The involution  (\ref{discr-inv}) and the commutation relations (\ref{cont-Heis})
 then imply the relations
\begin{gather}\label{inv-aa+}
(b_s^{\dagger})^*=b_s,
\\
\lbrack b_t,b_s^{\dagger}\rbrack=\delta(t-s), \qquad
\lbrack b_t^{\dagger},b_s^{\dagger}\rbrack=\lbrack b_t,b_s\rbrack=0,\nonumber
\end{gather}
where $\delta(t-s):= \delta_{s,t}$ is Kronecker's delta in the discrete
case and Dirac's delta in the continuous case.

The operator valued distribution form of the universal enveloping algebra ${\cal A}$,
of the Boson algebra, is the algebraic linear span of the expressions of the form
\[
b_{t_n}^{\varepsilon_n}\cdots b_{t_2}^{\varepsilon_2}b_{t_1}^{\varepsilon_1},
\]
where $n\in \mathbb{N}$, $t_1,\dots ,t_n\in\mathbb{R}$, $\varepsilon_j\in\{+,-\}$
and $b_{t_n}^{+}=b_t^{\dagger}$, $b_{t_n}^{-}=b_t$.

This has a natural structure of $*$-algebra induced by~(\ref{inv-aa+}).
On this algebra there is a par\-ticularly simple state characterized by the following theorem.
We outline a proof of this theorem because it illustrates in a simple case the path
we have followed to construct analogues of that state in much more complex
situations, namely:
\begin{enumerate}\itemsep=0pt
\item[$(i)$] to formulate an analogue of the Fock condition (\ref{df-Fk-st-A});

\item[$(ii)$] to use the commutation relations to associate a distribution kernel to
any linear functional~$\varphi$, satisfying the analogue of the Fock condition,
in such a way that $\varphi$ is positive if and only if
this kernel is positive def\/inite;

\item[$(iii)$] to prove that this kernel is ef\/fectively positive def\/inite.
\end{enumerate}

\begin{theorem}\label{ex-un-Fk-st}
On the $*$-algebra ${\cal A}$, defined above, there exists a unique
state $\varphi$ satisfying
\begin{equation}\label{df-Fk-st-A}
\varphi (b_t^{\dagger}x)= \varphi (xb_t) =0
, \qquad \forall \, x\in{\cal A}.
\end{equation}
\end{theorem}

\begin{proof}
From the commutation relations we know that ${\cal A}$ is
the algebraic linear span of expressions of the form
\begin{equation}\label{norm-ord}
b_{s_1}^{\dagger}b_{s_2}^{\dagger} \cdots b_{s_m}^{\dagger}
b_{t_n}\cdots b_{t_2}b_{t_1}
\end{equation}
 (normally ordered products), interpreted as the central element $1$ if both $m=n=0$.
Therefore, if a state $\varphi$ satisfying (\ref{df-Fk-st-A}) exists, then it is
uniquely def\/ined by the properties that $\varphi (1) = 1$ and $\varphi (x) = 0$
for any $x$ of the form (\ref{norm-ord}) with either $m$ or $n$ $\not=0$.
It remains to prove that the linear functional def\/ined by these properties is positive.

To this goal notice that the commutation relations imply that ${\cal A}$ is also
the algebraic linear span of expressions of the form
\[
b_{s_1}b_{s_2} \cdots b_{s_m}
b_{t_n}^{\dagger}\cdots b_{t_2}^{\dagger}b_{t_1}^{\dagger}
\]
(anti-normally ordered products), interpreted as before. A linear functional
$\varphi$ on ${\cal A}$ is positive if and only if the distribution kernel
\[
K\left( s_1, s_2,\dots ,s_m; t_1, t_2 ,\dots, t_2 ,t_n\right) :=
\varphi\left( b_{s_1}b_{s_2} \cdots b_{s_m}
b_{t_n}^{\dagger}\cdots b_{t_2}^{\dagger}b_{t_1}^{\dagger}\right)
\]
is positive def\/inite. If $\varphi$ satisf\/ies condition (\ref{df-Fk-st-A}), then
the above kernel is equal to (in obvious notations)
\begin{gather*}
\varphi\left( b_{s_1} \cdots b_{s_{m-1}}[b_{s_{m}},
b_{t_n}^{\dagger}\cdots b_{t_2}^{\dagger}b_{t_1}^{\dagger}]\right)=
\sum_{h=1}^{n-1}  \varphi\left( b_{s_1} \cdots b_{s_{m-1}}
b_{t_n}^{\dagger}\cdots b_{t_h+1}^{\dagger} [b_{s_{m}}, b_{t_h}^{\dagger}]b_{t_{h-1}}^{\dagger}
\cdots  b_{t_1}^{\dagger}\right)
\\
\phantom{\varphi\left( b_{s_1} \cdots b_{s_{m-1}}[b_{s_{m}},
b_{t_n}^{\dagger}\cdots b_{t_2}^{\dagger}b_{t_1}^{\dagger}]\right)}{}
=\sum_{h=1}^{n-1}  \delta(s_{m} - t_h)
\varphi\!\left( b_{s_1} \cdots b_{s_{m-1}}
b_{t_n}^{\dagger}\cdots b_{t_h+1}^{\dagger} b_{t_{h-1}}^{\dagger} \cdots  b_{t_1}^{\dagger}\right).
\end{gather*}
 From this, one deduces that the kernel
$K\left( s_1, s_2,\dots ,s_m; t_1, t_2 ,\dots, t_2 ,t_n\right)$
can be non zero if and only if $m=n$. Finally, since $ \delta(s - t)$ is a positive
def\/inite distribution kernel, the positivity of~$\varphi$ follows, by induction,
from Schur's lemma.
\end{proof}

\begin{definition}\label{df-Fock-st}
The unique state $\varphi$ on ${\cal A}$, def\/ined by Theorem~\ref{ex-un-Fk-st}
above, is called the Fock (or lowest weight) state.

The GNS representation $\{ {\cal H},\Phi,\pi \}$ of the pair $\{{\cal A},\varphi\}$
is characterized by:
\begin{equation}\label{equiv-char-Fk-st}
\pi(b_t)\Phi = 0
\end{equation}
in the operator valued distribution sense.
\end{definition}

\begin{definition} \label{df-Bos-Fock-wn}
In the notations of Def\/inition~\ref{df-Fock-st}
the operator (more precisely, the operator valued distribution) process
in $\{{\cal A},\varphi\}$, def\/ined by
\begin{equation*}
\{\pi(b_t^{\dagger}),\pi(b_t)\}
\end{equation*}
is called the {\it Boson Fock (or standard quantum) white noise} on $\mathbb{R}$.
\end{definition}

 Def\/inition \ref{df-Bos-Fock-wn} is motivated by the following theorem.

\begin{theorem}\label{th-stand-wn}
In the notation \eqref{df-aa+}, the two operator subprocesses in $\{{\cal A},\varphi\}$:
\begin{equation}\label{stand-wn}
\{\pi(q_t)\}  \qquad \hbox{and } \qquad \{\pi(p_t)\}
\end{equation}
  are classical processes stochastically isomorphic to the standard classical white noise
on $\mathbb{R}$.
\end{theorem}

\begin{proof} The idea of the proof is that the Fock state has clearly mean zero. Using a modif\/ication
of the argument used in the proof of Theorem~\ref{ex-un-Fk-st} one shows that
it is Gaussian and delta-correlated, i.e.\ it is by def\/inition a standard classical white noise.
\end{proof}

\begin{remark}
Theorem~\ref{th-stand-wn} and the relations~(\ref{df-aa+}) show that the
Boson Fock white noise is equivalent to the pair~(\ref{stand-wn}), of standard
classical white noises on $\mathbb{R}$. However the commutation relations~(\ref{cont-Heis}) show that these two classical white noises do not commute so that
classical probability does not determine their mixed moments:
this additional information is provided by quantum probability.
\end{remark}

 In the following, when no confusion can arise, we omit from the notations the symbol $\pi$ of the representation.

\section[Current algebras over $\mathbb{R}^d$]{Current algebras over $\boldsymbol{\mathbb{R}^d}$}\label{curr-alg}

Current algebras are associated to pairs:
($*$-Lie algebra, set of generators) as follows.
Let ${\cal G}$ be a $*$-Lie algebra with a set of generators
\[
\{ l^{+}_{\alpha}, l^{-}_{\alpha},  l^{0}_{\beta}   :  \ \alpha\in I , \ \beta\in I_0 \},
\]
 where $I$, $I_0$ are sets satisfying
\[
I \cap I_0 = \varnothing
\]
and $(c^\gamma_{\alpha\beta}(\varepsilon,\varepsilon',\varepsilon''))$ are the structure constants corresponding to the given set of generators, so that:
\begin{equation}\label{str-const}
[l^\varepsilon_\alpha,l^{\varepsilon'}_\beta]=
c^\gamma_{\alpha\beta}(\varepsilon,\varepsilon',\varepsilon'')
l^{\varepsilon''}_\gamma.
\end{equation}
Here and in the following, summation over repeated indices is understood.
The sets $I$, $I_0$ can, and in the examples below will, be inf\/inite. However, here and in
the following, we require that, in the summation on the right hand side of~(\ref{str-const}), only a f\/inite number of terms are
non-zero or equivalently that the structure constants are almost all zero
(also this condition is automatically satisf\/ied in the examples below). We assume that
\[
(l^{+}_{\alpha})^* = l^{-}_{\alpha}  , \qquad \forall\, \alpha\in I
, \qquad (l^{0}_{\beta})^* = l^{0}_{\beta}  , \qquad  \forall\, \beta\in I_0.
\]

The transition to the current algebra of ${\cal G}$ over $\mathbb{R}^d$ is obtained by replacing the generators by ${\cal G}$-valued distributions on $\mathbb{R}^d$
\[
l^\varepsilon_\alpha\to l^\varepsilon_\alpha(x), \qquad
x\in \mathbb{R}^d
\] and the corresponding relations by
\begin{gather*}
l^{+}_{\alpha}(x)^*=l^{-}_{\alpha}(x), \qquad l^{0}_{\beta}(x)^*=l^{0}_{\beta}(x)
, \qquad  \forall\, \alpha\in I  , \qquad \forall \, \beta\in I_0,
\\
[l^\varepsilon_\alpha(x),l^{\varepsilon'}_\beta(y)]=
c^\gamma_{\alpha\beta}(\varepsilon,\varepsilon',\varepsilon'') l^{\varepsilon''}_\gamma(x)
\delta(x-y).
\end{gather*}

 This means that the structure constants are replaced by
\[
c^\gamma_{\alpha\beta}(\varepsilon,\varepsilon',\varepsilon'')
\to
c^\gamma_{\alpha\beta}(\varepsilon,\varepsilon',\varepsilon'') \delta(x-y).
\]
In terms of test functions this can be equivalently formulated as follows.

\begin{definition} Let ${\cal G}$ be a $*$-Lie algebra with generators
\[
\{ l^{+}_{\alpha}, l^{-}_{\alpha},  l^{0}_{\beta}  : \ \alpha\in I  , \ \beta\in I_0 \}
\]
and let ${\cal C}$ be a vector space of functions from
$\mathbb{R}^d$ to $\mathbb{C}$ called the test function space.
A current algebra of ${\cal G}$ over $\mathbb{R}^d$ with test function space
${\cal C}$ is a $*$-Lie algebra with generators
\[
\{ l^{+}_{\alpha}(f), l^{-}_{\alpha}(g),  l^{0}_{\beta}(h)   : \
\alpha\in I , \ \beta\in I_0  , \ f,g,h\in   {\cal C}\}
\]
such that the maps
\[
f\in   {\cal G}\mapsto l^{+}_{\alpha}(f)  , \ l^{0}_{\beta}(f)
\]
are complex linear, the involution satisf\/ies
\[
(l^{+}_{\alpha}(f))^* = l^{-}_{\alpha}(f), \qquad
(l^{0}_{\beta}(f))^* =  l^{0}_{\beta}(\bar f), \qquad
\forall\, \alpha\in I   , \qquad  \forall\, \beta\in I_0   , \qquad \forall \, f\in   {\cal C}
\]
and the commutation relations are given by:
\[
[l^\varepsilon_\alpha(f),l^{\varepsilon'}_\beta(g)] :=
c^\gamma_{\alpha\beta}(\varepsilon,\varepsilon',\varepsilon'')
l^{\varepsilon''}_\gamma(f^\varepsilon g^{\varepsilon'})
\]
with the convention:
\[
f^\varepsilon = \begin{cases}
\overline f,\quad\hbox{if}\quad \varepsilon = -,\\
f ,\quad\hbox{if}\quad \varepsilon\in\{0,+\}.
 \end{cases}
\]
\end{definition}

\begin{remark}
By restriction of the test function space ${\cal C}$ to real valued functions, or more ge\-ne\-rally by restricting one's attention to
real Lie algebras, one could avoid the introduction of generators
and use an intrinsic def\/inition. We have chosen the non-intrinsic formulation because we want to emphasize the intuitive analogy between the generators $l^\pm_\alpha$ and the powers of the creation/annihilation operators and between the generators $l^0_\alpha$ and the powers of the number operator. Thus, for example, in the RHPWN $*$-Lie algebra with generators $B^n_k$, the indices $\alpha\in I$ are the pairs $(n,k)$ with $n>k$ and the indices $\beta\in I_0$ are the diagonal pairs $(n,n)$.
\end{remark}

\section{Connection with independent increment processes}\label{Prob-impl}

Let ${\cal G}({\cal C})$ be Lie algebra whose elements depend
on test functions belonging to a certain space~${\cal C}$
of functions $f: \Bbb R \rightarrow \Bbb C$.
Then ${\cal G}$ has a natural localization, obtained by f\/ixing
a family ${\cal F}$ of subsets of~$\Bbb R$, e.g.\ intervals,
and def\/ining the subalgebra
\[
{\cal G}_I:=\{L(f)\in {\cal G}:\  \hbox{supp}(f) \subseteq I\}.
\]
Suppose that ${\cal G}$ enjoys the following property:
$\forall \, I, J \subseteq \Bbb R$
\[
I\cap J=\varnothing \ \Rightarrow \ [{\cal G}_I,{\cal G}_J]=0
\]
(notice that, if ${\cal G}({\cal C})$ is a current algebra over $\Bbb R$ of a Lie algebra
${\cal G}$, then this property is automatically satisf\/ied).
If this is the case, denoting ${\cal A}_I$ the algebraic linear span of
(the image of)~${\cal G}_I$ in any representation, the (associative)
$*$-algebra generated by ${\cal A}_I$
and ${\cal A}_J$ is the linear span of the products of the form $a_Ia_J$ where $a_I$
(resp.~$a_J$) is in  ${\cal A}_I$ (resp.~${\cal A}_J$).

A similar conclusion holds if, instead of two disjoint sets, one considers an
arbitrary f\/inite number of disjoint sets.
Denote ${\cal A}$ the algebraic linear span of~${\cal G}$ in a representation
with cyclic vector $\Phi$ and $\varphi$, the restriction of the state
$\langle \Phi, ( \, \cdot \, ) \Phi \rangle$ to ${\cal A}$.
The given cyclic representation and the state  $\varphi$
are called {\it factorizable} if for any f\/inite family $I_1,\dots,I_n$,
of mutually disjoint intervals of~$\Bbb R$ one has{\samepage
\[
\varphi(a_{I_1}\dots a_{I_n})= \prod^n_{j=1} \varphi(a_{I_j})
, \qquad a_{I_j} \in {\cal A}_{I_j}, \qquad j\in\{1,\dots,n\}.
\]
The Fock representation, and all its generalizations we have
considered so far, have this property.}

By restriction to Abelian subalgebras, factorizable representations
give rise to classical (polynomially) independent increment processes.
The above def\/inition of factorizability applies to general linear functionals
(i.e.\ not necessarily positive or normalized). In the following we will
make use of this remark.

Given a cyclic representation $\{{\cal H},\pi,\Phi\}$ of ${\cal G}$
(we omit $\pi$ from notations), for any interval $I$ one def\/ines the subspace ${\cal H}_I$
of ${\cal H}$ as the closed subspace containing $\Phi$ and invariant
under ${\cal G}_I$.

\section{Quadratic powers: brief historical survey}\label{sec-quadr-pow}

The commutation relations imply that
\[
[b^{2}_s , b^{+2}_t] = 4\delta(t -s )b^{+}_s b_t + 2\delta(t -s )^2
\]
and the appearance of the term $\delta(t -s )^2$ shows that
$b^{+2}_t$ and $b^2_t$ are not well def\/ined even as operator valued distributions.
The following formula, due to Ivanov, for the square of the delta function
(cf.~\cite{[Ivanov79]} for a discussion of its precise meaning)
\begin{equation}\label{Ivan-form}
\delta^2(t)=c\,\delta(t),\qquad \hbox{$c$ is arbitrary constant},
\end{equation}
was used by Accardi, Lu and Volovich to realize the program discussed
in Section~\ref{sec-HPWN-schm} for the second powers of~WN.

Using this we f\/ind the renormalized commutation relation:
\begin{equation}\label{quadr-com-rel-BB+}
[b^{2}_s , b^{+2}_t] = 4\delta(t -s )b^{+}_s b_t + 2c\delta(t -s ).
\end{equation}
Moreover (without any renormalization!)
\begin{equation}\label{quadr-com-rel-BN}
[b^{2}_s ,  b^{+}_tb_t] = 2\delta(t -s )b^{2}_t.
\end{equation}
Introducing a test function space (e.g.\ the complex valued step functions on
$\mathbb{R}$ with f\/initely many values), one verif\/ies that the smeared operators
(see the comments at the beginning of Section~\ref{sec-HPWN-schm} about their meaning)
\begin{equation}\label{quadr-ops-tf}
b^+_\varphi =\int dt\varphi(t)b^2_t,\qquad b_\varphi= (b^+_\varphi)^+
,\qquad
n_\varphi^* = n_\varphi =\int dt\varphi(t)b^+_tb_t
\end{equation}
satisfy the commutation relations
\begin{gather*}
[b_\varphi,b^+_\psi]=c\langle\varphi,\psi\rangle+n_{\overline\varphi\psi},
\qquad
[n_\varphi,b_\psi]=-2b_{\overline\varphi\psi},
\qquad
[n_\varphi,b^+_\psi]=2b^+_{\varphi\psi},\\
(b^+_\varphi)^+=b_\varphi, \qquad n^+_\varphi=n_{\overline\varphi}.
\end{gather*}
The relations (\ref{quadr-com-rel-BB+}), (\ref{quadr-com-rel-BN}), or their
equivalent formulation in terms of the generators (\ref{quadr-ops-tf}),
are then taken as the def\/inition of the renormalized square of white noise
(RSWN) $*$-Lie algebra.
Recalling that $sl(2,  \mathbb{R})$ is the $*$-Lie algebra with 3 generators
$B^-, \ B^+ , \ M $ and relations
\begin{gather*}
[B^-,B^+]=M, \qquad
[M,B^\pm]=\pm2B^\pm, \qquad
(B^-)^* =B^+, \qquad
M^* =M
\end{gather*}
one concludes that the RSWN $*$-Lie algebra is isomorphic to a current algebra, over
$\mathbb{R}$, of a~central extension of $sl(2,  \mathbb{R})$. Notice that this
central extension is trivial (like all those of~$sl(2,  \mathbb{R})$), but its role
is essential because without it, i.e.\ putting $c=0$ in the commutation relations
(\ref{quadr-com-rel-BB+}), the Fock representation discussed below
reduces to the zero representation.

Keeping in mind the intuitive expressions (\ref{quadr-ops-tf}) of the generators,
a natural analogue of the characterizing property (\ref{df-Fk-st-A}), of the Fock
state for this algebra, would be
\begin{equation}\label{df-quad-Fk-st-A}
\varphi (b_t^{\dagger2}x)=
\varphi (xb_t^{2}) =\varphi (b_t^{\dagger}b_tx)= \varphi (xb_t^{\dagger}b_t)
,\qquad  \forall \, x\in U(sl(2,  \mathbb{R}))
\end{equation}
(let us emphasize that (\ref{df-quad-Fk-st-A}) is only in an informal sense a particular
case of (\ref{df-Fk-st-A}) where diagonal terms were not included) or, using test
functions and the equivalent characterization~(\ref{equiv-char-Fk-st}) of the f\/irst order Fock state:
\[
b_\varphi \Phi =  n_\varphi \Phi =0.
\]
Using this property as the def\/inition of the {\it quadratic Fock state},
Accardi, Lu and Volovich proved in \cite{[AcLuVo99]} the existence of
the quadratic Fock representation $ \{ {\cal H} , \Phi , \pi\}  $
and formulated the programme to achieve a similar result for higher powers,
using a natural generalization of the renormalization used for the square
(see Section~\ref{sec-Iv-ren-HP}).

The paper \cite{[AcLuVo99]} opened a research programme leading to
several investigations in dif\/ferent directions. Among them we mention below only
those directly related to the representation theory of Lie algebras and we refer,
for more analytical and probabilistic directions, to~\cite{[AcBou05g],[AcBou06a],[AcBou01e],[AcBou01d],[AcBou03a],[AcBou01f],[AcHiKu01],[AcRo05]}. The latter paper also includes a discussion of previous
attempts to give a meaning to the squares of free f\/ields.

 $(i)$  Accardi and Skeide  introduced in \cite{[AcSk99b]} the quadratic exponential (coherent) vectors for the RSWN and noticed that the kernel def\/ined by the scalar product of two such vectors coincided with the kernel used by Boukas and Feinsilver in \cite{Bou91,pfein} and \cite{Fein}  to construct unitary representations of the so-called  Finite Dif\/ference Lie Algebra. Moreover, they proved that the Fock representation of the RSWN $*$-Lie algebra, constructed in~\cite{[AcLuVo99]}, gave rise to a type-I product system of Hilbert spaces in the sense of Arveson (cf.~\cite{[Arv03]}).

 $(ii)$ Accardi, Franz and Skeide  realized in
\cite{[AcFrSk00]} that the RSWN $*$-Lie algebra is a current algebra of $sl(2,\mathbb{R})$ over $\mathbb{R}$ and that the factorization property mentioned in item $(i)$ above naturally suggested a connection with the theory of inf\/initely divisible stochastic processes along the lines described in the monographs~\cite{Guichardet} and \cite{Partha-Schmidt}. In particular they were able to identify the inf\/initely divisible classical stochastic
processes, arising as vacuum distributions of the generalized f\/ield operators of the RSWN, with the three non-standard classes of Meixner laws:
\begin{enumerate}\itemsep=0pt
\item[--] Gamma,
\item[--] Negative binomial (or Pascal),
\item[--] Meixner.
\end{enumerate}

Since it was well known that the remaining two classes of Meixner laws, i.e.\ the Gaussian and Poisson classes, arise as vacuum distributions of the generalized f\/ield operators of the usual f\/irst order white noise (free boson f\/ield), this result showed that the quantum probabilistic approach provided a nice unif\/ied view to the 5 Meixner classes which were discovered in 1934 (cf.~\cite{[Meix34]}) in connection with a completely dif\/ferent problem (a survey of this development is contained in~\cite{[AcBou04c]}).

For a concrete example on how some Meixner laws can appear as
vacuum distributions of quantum observables, see Section~\ref{Cl-stoch-pr-gen-Fk-rep-RHPWN} below.

 $(iii)$ P.~Sniady  constructed in \cite{[Snia99]} the free analogue the Fock representation of the RSWN $*$-Lie algebra obtained in~\cite{[AcLuVo99]} and proved the f\/irst no-go theorem concerning the impossibility of combining together in a nontrivial way the Fock representations of the f\/irst and second order white noise $*$-Lie algebras. This opened the way to a series of no-go theorems which paralleled, in a quite dif\/ferent context and using dif\/ferent techniques, a series of such theorems obtained in the physical literature.

 A stronger form of Sniady's result, still dealing with the f\/irst and second order case, was later obtained in~\cite{[AcFrSk00]}; in
\cite{[AcBouFr06]} this result was extended to the higher powers, def\/ined with the renormalization used in~\cite{[AcLuVo99]}, and further extended to the higher powers of the $q$-deformed white noise \cite{[AcBou05a]}.

 $(iv)$ The attempt to go beyond the Fock representation  by constructing more
general representations, such as the f\/inite temperature one, related to KMS states,
was initiated in~\cite{[AcAmFr02]} where the analogue of the Bogolyubov transformations for the RSWN was introduced (i.e.\ those transformations on the test
function space which induce endomorphisms of the quadratic $*$-Lie algebra) and
a (very particular) class of KMS states on the RSWN algebra was constructed.

The problem of constructing the most general KMS states (for the free
quadratic evolution) on the RSWN algebra was attacked with algebraic techniques
in the paper~\cite{[AcPeRo04]} but its solution was obtained later,
with a purely analytical approach by Prohorenko~\cite{[Prohor05]}.

$(v)$ The quadratic Fermi case was investigated by Accardi, Arefeva and Volovich
in \cite{[AcArVo03]} and led to the rather surprising conclusion that, while the
quadratic Bose case leads to the representation theory of the compact form of
the real Lie algebra $SL(2,\Bbb R)$, the corresponding Fermi case leads to
the non compact form of the same real Lie algebra.

\section{Higher powers of white noise}\label{sec-HPWN-schm}

In order to realize, for the higher powers of white noise, what has been achieved
for the square, we def\/ine the smeared operators (we will often use this terminology
which can be justif\/ied only a posteriori by the realization of these objects as
linear operators on Hilbert spaces):
\begin{equation}\label{smrd-op-HP}
B_k^n(f):=\int_{\mathbb{R}} f(t) {b_t^{\dagger}}^n b_t^k dt.
\end{equation}
Notice that the above integral is normally ordered in $b_t^{\dagger}$, $b_t$
therefore it always has a meaning as a~sesquilinear form on the (f\/irst order)
exponential vectors with test function in $L^1\cap L^{\infty}(\mathbb{R})$
independently of any renormalization rule.
This allows to consider the symbols $B_k^n(f)$ as generators of a complex vector
space (in fact a $*$-vector space) and also to introduce a topology.

It is only when we want to introduce an additional Lie
algebra structure, which {\it keeps some track} of the discrete version of the symbolic
commutation relations written below (see formula~(\ref{HO-CR})),
that a renormalization rule is needed.

In terms of creation and annihilation operators, the commutation relations \eqref{cont-HOCCR}
take the form (see \cite{[AcBouFr06]} for a proof):
\begin{gather}\label{HO-CR}
\lbrack{b_t^{\dagger}}^nb_t^k,{b_s^{\dagger}}^Nb_s^K\rbrack
\\
\qquad= \sum_{L\geq 1}
\binom{k}{L}N^{(L)}  \left\{
\epsilon_{k,0} \epsilon_{N,0}{b_t^{\dagger}}^{n} {b_s^{\dagger}}^{N-L}b_t^{k-L} b_s^K
- [ (k,n) \leftrightarrow (K,N)]\right\} \delta^L(t-s),\nonumber
\end{gather}
where $n,k,N,K\in\mathbb{N}$, $x^{(y)}$ is the Pochammer symbol def\/ined
in formula (\ref{Poch-symb}), $[ (k,n) \leftrightarrow (K,N)]$ denotes the result
obtained by exchanging the roles of  $(k,n)$ and $(K,N)$ in the expression $\epsilon_{k,0}\,\epsilon_{N,0}{b_t^{\dagger}}^{n}\,{b_s^{\dagger}}^{N-L}\,b_t^{k-L}\,b_s^K$ and, by def\/inition:
\[
 \epsilon_{n,k}:=1-\delta_{n,k}  \qquad \hbox{(Kroneker's delta)}.
\]
In order to give a meaning to the powers $\delta^L(t-s)$, of Dirac's delta, for $L\geq 2$,
we f\/ix a~renormalization rule and a space of test functions ${\cal T}$.
After that, from the distribution-form of the commutation relations (\ref{HO-CR}),
one deduces the corresponding commutation rules and involution for these operators.

Notice that it is not {\em a priori} obvious that, after the modif\/ications
introduced by the renormalization rule, the resulting brackets still def\/ine
a $*$-Lie algebra. This fact has to be checked case by case.
The following section gives a f\/irst illustration of the procedure described above.

\subsection{Renormalization based on Ivanov's formula}\label{sec-Iv-ren-HP}

The obvious generalization of Ivanov's formula (\ref{Ivan-form})
to powers strictly higher than two
\[
\delta^l(t)=c^{l-1} \delta(t),\qquad l=2,3,\dots, \qquad c>0
\] leads to the following commutation relations and involution:
\begin{gather}\label{HO-CR-Iv-ren}
\lbrack{b_t^{\dagger}}^nb_t^k,{b_s^{\dagger}}^Nb_s^K\rbrack
\\
\qquad{}= \sum_{L\geq 1} \binom{k}{L}N^{(L)}\, \left\{\epsilon_{k,0}\,\epsilon_{N,0}
{b_t^{\dagger}}^{n}\,{b_s^{\dagger}}^{N-L}\,b_t^{k-L}\,b_s^K- \ [ (k,n)
\leftrightarrow (K,N)]\right\}\,c^{L - 1}\,\delta (t-s)\nonumber
\end{gather}
 or equivalently in terms of test functions
\begin{gather}\label{RHO-CR1}
\lbrack B^N_K( g),B^n_k(f) \rbrack=
\sum_{L= 1}^{ (K \wedge n) \vee ( k \wedge N)  }
\theta_L (N,K;n,k)
 c^{L - 1} B^{N + n - L}_{K + k - L} ( g f),
\\
\label{RHO-inv1}
 B^N_K( \bar g)^* =  B^K_N( \bar g),
\end{gather}
where by def\/inition:
\begin{gather*}
\theta_L(N,K;n,k):=H(L-1)  \left(\epsilon_{K, 0}\,\epsilon_{n, 0} \binom{K,n}{ L} - \epsilon_{k, 0} \epsilon_{N,0}  \binom{k,N}{L} \right),
\\
\binom{y,z}{x}:=\binom{y}{x}\,z^{(x)},\qquad H(x)=\left\{
\begin{array}{ll}
1&\mbox{ if } x \geq 0,\\
0&\mbox{ if } x < 0.
\end{array}
\right.
\end{gather*}

One can prove that the brackets and involution (\ref{RHO-CR1}) and
(\ref{RHO-inv1}) ef\/fectively def\/ine a $*$-Lie algebra: the renormalized higher
powers of white noise (RHPWN) $*$-Lie algebra. This can be proved directly, but
the simplest proof is based on the remark that an inspection of formula~(\ref{HO-CR-Iv-ren}) shows that it coincides with the
prescription to construct a current algebra over $\mathbb{R}$ of the universal
enveloping algebra of the $1$-dimensional Heisenberg algebra discussed
in Section~\ref{sec-1-d-cse} and denoted~${\cal A}(\Bbb R)$ (cf.~\cite{[AcBo07]}).

In the following section
we discuss the notion of current algebra in some generality because, in the probabilistic interpretation of $*$-representations of Lie algebras, the transition from a Lie
algebra to an associated current algebras corresponds to the transition from a random variable to a stochastic process (or random f\/ield).

\subsection[Fock representation for RHPWN defined using Ivanov's renormalization
and corresponding no-go theorems]{Fock representation for RHPWN def\/ined using Ivanov's renormalization
and corresponding no-go theorems}\label{nogo-thm-Iv-nrm}

In this section the test function space will be the space of complex valued
step functions on $\mathbb{R}$ with f\/initely many values. Using the linear (or
anti-linear) dependence of the generators on the test functions, one can restrict  to characteristic functions $\chi_I$ of intervals  $I\subseteq \mathbb{R}$
(i.e.\ $\chi_I(x)=1$ if $x\in I$, $\chi_I(x)=0$ if $x\notin I$)
and often we simply write
\begin{equation}\label{charfunc}
B^{n}_k := B^{n}_k(\chi_I).
\end{equation}
Given the expression (\ref{smrd-op-HP}) of the generators $B^n_k(f)$
a natural way to extend to them the notion of Fock representation is the following:

\begin{definition}\label{df-Fk-rep-RHPWN1}
A cyclic representation $\{ {\cal H}, \pi, \Phi\}$ of the RHPWN $*$-Lie algebra
is called Fock if (omitting as usual $\pi$ from the notations)
\begin{equation}\label{df-Fk-RHPWN-Iv}
B^n_k(f)\Phi = 0 , \qquad  \forall \, f , \qquad  \forall\,  k\geq n.
\end{equation}
\end{definition}

 One can prove that condition (\ref{df-Fk-RHPWN-Iv}) ef\/fectively def\/ines
a linear functional on the universal enveloping algebra of the RHPWN $*$-Lie algebra
and that this functional is factorizable. This allows to restrict the proof of positivity
to  the $*$-algebra generated by a single operator of the form (\ref{charfunc}).

 The obstructions to the positivity requirements are illustrated by
the following no-go theorem.

\begin{theorem}[no-go theorem \cite{[AcBouFr06]}]\label{nogo-thm1}
Let $\mathcal{L}$ be a Lie $*$-subalgebra of RHPWN. Suppose that
\begin{enumerate}\itemsep=0pt
\item[$(i)$] for some $n\geq 1$, $\mathcal{L}$ contains
$B^n_0$ $(n$-th creator power$)$ and $B^{2n}_0$,
\item[$(ii)$] the test function space includes functions whose support has Lebesgue measure smaller than~$1/c$ $($the inverse of the renormalization constant$)$.
    \end{enumerate}
 Then $\mathcal{L}$ does not admit a Fock representation.
\end{theorem}

\begin{proof} The idea of the proof is the following. Assuming that a Fock representation exists and using the above assumptions, one constructs negative norm vectors (ghosts) by considering linear combinations of $B^n_0\Phi$ and $B^{2n}_0\Phi$.
\end{proof}

\begin{corollary}\label{Bn-B2n}
In the notations of Theorem~{\rm \ref{nogo-thm1}}, if $\mathcal{L}$ contains $B^3_0$
then it does not admit a Fock representation.
\end{corollary}

\begin{proof} The idea of the proof is that, if $\mathcal{L}$ contains $B^3_0$, then its cyclic space must contain also $B^{6}_0\Phi$ and then applies Theorem~\ref{nogo-thm1}.
\end{proof}

\begin{corollary}\label{Schr-alg}
The current algebra over $\mathbb{R}$ of the Schr\"odinger algebra, i.e.\ the $*$-Lie
algebra with generators
\[
\{ a^{+}, a , a^{+2}, a^2, a^{+}a , 1\}
\] does not admit a Fock representation if the test function space
includes functions whose support has Lebesgue measure smaller than the inverse
of the renormalization constant.
\end{corollary}

\begin{proof}  The idea of the proof is that the Schr\"odinger algebra contains $a^{+}$ and $a^{+2}$. Therefore the associated current algebra over $\mathbb{R}$ contains $B^{1}_0(\chi_I)$
and $B^{2}_0(\chi_I)$ for arbitrary small intervals $I\subseteq \mathbb{R}$. The thesis then follows from Theorem \ref{nogo-thm1}.
\end{proof}

\section{A new renormalization}\label{section9}

The no-go theorems, mentioned in Section~\ref{nogo-thm-Iv-nrm}
above, emphasize the necessity to investigate other renormalization procedures in
order to go beyond the square and  construct explicitly the (or better a) $*$-Lie algebra
canonically associated with the renormalized higher powers of white noise.

In the attempt to overcome the no-go theorems, Accardi and Boukas
introduced another, convolution type, renormalization of $\delta^l(t)$:
\begin{equation}\label{conv-ren}
\delta^l(t-s)=\delta(s)\,\delta(t-s), \qquad l=2,3,\dots.
\end{equation}
We refer to \cite{[AcBo06]} for the motivations which led to this special choice. The involution is the same as in (\ref{RHO-inv1}) while the commutation relations resulting from the renormalization prescription (\ref{conv-ren}) are:
\begin{gather*}
[{b_t^{\dagger}}^nb_t^k,{b_s^{\dagger}}^Nb_s^K] 
\\
\qquad{}=
\epsilon_{k,0}\epsilon_{N,0} ( k N  {b_t^{\dagger}}^{n} {b_s^{\dagger}}^{N-1} b_t^{k-1} b_s^k \delta(t-s)
 +    \sum_{L\geq 2} \binom{k}{L}N^{(L)} {b_t^{\dagger}}^{n} {b_s^{\dagger}}^{N-L} b_t^{k-L} b_s^K  \delta(s) \delta(t-s) )
\nonumber\\
\qquad{}
-\epsilon_{K,0}\epsilon_{n,0}( K n  {b_s^{\dagger}}^{N} {b_t^{\dagger}}^{n-1} b_s^{K-1} b_t^k \delta(t-s)
  +   \sum_{L\geq 2} \binom{K}{L}n^{(L)} {b_s^{\dagger}}^{N} {b_t^{\dagger}}^{n-L} b_s^{K-L} b_t^k \delta(s) \delta(t-s)),
  \nonumber
\end{gather*}
which, after  multiplying both sides  by  $f(t) g (s)$ and  integrating the resulting identity, yield the commutation relations
\begin{gather*}
\lbrack B^n_k( g),B^N_K(f) \rbrack=\left(\epsilon_{k,0} \epsilon_{N,0}  k N- \epsilon_{K,0} \epsilon_{n,0}  K n \right)  B^{N+n-1}_{K+k-1}( g f)
\\
\phantom{\lbrack B^n_k( g),B^N_K(f) \rbrack=}{} +\sum_{L= 2}^{ (K \wedge n) \vee ( k \wedge N)  }  \theta_L (n,k;N,K) \bar g(0)  f(0) {b_0^{\dagger}}^{N+n-l} b_0^{K+k-l},
\end{gather*}
where $\theta_L (n,k;N,K) $ is as in (\ref{RHO-CR1}). By restricting the test function space to functions $f$, $g$ that satisfy the boundary condition
\begin{equation}\label{bdry-cond-0}
f(0)=g(0)=0
\end{equation}
we eliminate the singular terms ${b_0^{\dagger}}^{N+n-l} b_0^{K+k-l} $. The resulting commutation relations are given in the following def\/inition.

\begin{definition}
The RHPWN commutation relations are:
\begin{equation}\label{WN-CR}
\lbrack B^n_k( g),B^N_K(f) \rbrack_{\rm RHPWN}:=
\left( k  N- K   n  \right)  B^{n+N-1}_{k+K-1}( g f).
\end{equation}
\end{definition}

 These commutation relations exhibit a striking similarity to those of the $w_{\infty}$ Lie algebra with generators $\hat{B}^n_k$, arising in conformal f\/ield theory (cf.~\cite{6,8,9} and \cite{BK91}):
\begin{equation}\label{w-inf-CR}
\lbrack \hat{B}^n_k , \hat{B}^N_K \rbrack_{w_{\infty} } =\left(   k (N-1)-K\,(n-1)  \right)  \hat{B}^{n+N-2}_{k+K}.
\end{equation}

However, at odds with what happens in the RHPWN algebra, here $n,k\in\mathbb{Z}$, with $n\geq 2$, and the involution is given by
\begin{equation*}
\big( \hat{B}^n_k \big)^*= \hat{B}^n_{-k}.
\end{equation*}
In particular, for $n=2$, one f\/inds the centerless Virasoro (or Witt) Lie algebra commutation relations
\begin{equation*}
\lbrack\hat{B}^2_k,\hat{B}^2_{K} \rbrack_{Vir}:=(k-K) \hat{B}^{2}_{k+K}.
\end{equation*}

Even with these dif\/ferences, the similarity of the commutation relations
(\ref{WN-CR}) and~(\ref{w-inf-CR}), was too strong to be a chance.
This motivated several papers attempting to prove the identity  of the two algebras
(cf.~\cite{id}). The basic idea to identify the two algebras arose from an analysis
of their classical realizations in terms of Poisson brackets which is outlined in the following section.

\section[Classical representations of the RHPWN and $w_{\infty}$ Lie algebras]{Classical representations of the RHPWN\\ and $\boldsymbol{w_{\infty}}$ Lie algebras}\label{RHPWN-winf}

We say that a Lie algebra ${\cal G}$ with generators $(l_{\alpha})_{\alpha\in F}$
($F$ a set) and structure constants $(c^{\gamma}_{\alpha,\beta})$ (with the properties specif\/ied in Section~\ref{curr-alg}) admits a classical representation if there exists a~space of functions $\hat {\cal G}$ from some $\mathbb{R}^d$ (with even~$d$) and with values
in $\mathbb{C}$ such that
\begin{enumerate}\itemsep=0pt
\item[$(i)$] $\hat {\cal G}$, as a linear space, has a set of algebraic generators
$(\hat l_{\alpha})_{\alpha\in F}$ (i.e.\ any element of $\hat {\cal G}$ is a~linear combination of a f\/inite subset of the $(\hat l_{\alpha})_{\alpha\in F}$),

\item[$(ii)$] $\hat {\cal G}$ is a Lie algebra with brackets given by the Poisson-brackets:
\begin{equation*}
[f,g]_{\hat {\cal G}} := \frac{\hbar}{i} \{f,g\}=\frac{\hbar}{i} \left(
\frac{ \partial f }{  \partial x} \cdot \frac{ \partial g }{ \partial  y}-
\frac{ \partial f }{ \partial y} \cdot  \frac{ \partial g }{  \partial x }\right),
\end{equation*}
\item[$(iii)$] the structure constants of $\hat {\cal G}$ in the basis $(\hat l_{\alpha})_{\alpha\in F}$
are the $(c^{\gamma}_{\alpha,\beta})$, i.e.
\[
[\hat l_{\alpha}, \hat l_{\beta}]_{\hat {\cal G}} = c^{\gamma}_{\alpha,\beta}\hat l_{\gamma},
\]
equivalently: the map $l_{\alpha}\mapsto \hat l_{\alpha}$ extends to a Lie algebra isomorphism.
\end{enumerate}

 The following classical representation of the $w_{\infty}$ Lie algebra
was known in the literature (cf.~\cite{4a})
\begin{equation}\label{cl-re-winf}
\hat w_{\infty} := \hbox{linear span of }
\big\{f_{n,k}(x,y):=e^{ikx} y^{n-1}   : \ n,k \in \mathbb{Z} , \  n\geq 2 , \ x,y\in\mathbb{R}\big\}.
\end{equation}
In fact one verif\/ies that:
\[
\{f_{n,k}(x,y),f_{N,K}(x,y)\}=i \left(k(N-1)-K(n-1) \right)  f_{n+N-2,k+K}(x,y).
\]
Notice that, for $n=N=2$ one recovers a classical representation,
in the sense def\/ined at the beginning of the present section, of
the Witt--Virasoro algebra, in which the space $\hat {\cal G}$
is a space of trigonometric polynomials in two real variables. The usual
realization of the Witt--Virasoro algebra is in terms of vector f\/ields
on the unit circle.

The analogue classical representation of the RHPWN Lie algebra
was introduced in the pa\-pers~\cite{[AcBou06]} and \cite{[AcBo06]}
\begin{gather}\label{cl-re-WN}
\widehat {\rm RHPWN} := \hbox{linear span of } \Bigg\{
g_{n,k}:=\left(\frac{x+iy}{\sqrt{2}}\right)^n\left(\frac{x-iy}{ \sqrt{2}}\right)^k
  : \, n,k \in \mathbb{N}, \, x,y\in\mathbb{R} \Bigg\}.
\end{gather}
In fact one verif\/ies that:
\[
\{g_{n,k}(x,y),g_{N,K}(x,y)\}=i \left( kN-nK \right)  g_{n+N-1,k+K-1}(x,y).
\]

\section[White noise form of the $w_{\infty}$ generators]{White noise form of the $\boldsymbol{w_{\infty}}$ generators}\label{section11}

Comparing (\ref{cl-re-winf}) and (\ref{cl-re-WN}) one realizes that,
although the two algebras are dif\/ferent (because the coef\/f\/icients of one are
generated by monomials and those of the other by trigonometric polynomials),
their closures in many natural topologies are the same.
Therefore it is natural to conjecture that a similar relationship holds
also in the quantum case.

After some guessing and corresponding trial and error attempts
the following (quantum and continuum) generalization of
(\ref{cl-re-winf}) was established in \cite{id}
\begin{equation}\label{wn-rep-of-winf}
\hat{B}_k^n(f):=\int_{\mathbb{R}} f(t) e^{ \frac{k}{2}(b_t- b_t^{\dagger})}
\left(\frac{ b_t+ b_t^{\dagger}}{2}\right)^{n-1}    e^{ \frac{k}{2}(b_t- b_t^{\dagger})} dt.
\end{equation}
For $n=2$ one obtains the Boson representation of
the centerless Virasoro (or Witt) Lie algebra generators
\[
\hat{B}_k^2(f):=\int_{\mathbb{R}} f(t) e^{ \frac{k}{2}(b_t- b_t^{\dagger})}
\left(\frac{ b_t+ b_t^{\dagger}}{2}\right)   e^{ \frac{k}{2}(b_t- b_t^{\dagger})} dt.
\] Expressions like (\ref{wn-rep-of-winf}) are symbolic expressions which involve ill
def\/ined quantities such as exponentials and products of operator valued distribution.
In order to give them a precise meaning one adopts the usual strategy in the theory
of distributions: the symbols are manipulated by formally applying to them the
f\/irst order commutation relations, then applying the renorma\-li\-zation prescription
(\ref{conv-ren}) and f\/inally integrating over test functions which satisfy
the boundary condition (\ref{bdry-cond-0}).

After having played this game with (\ref{wn-rep-of-winf}), one arrives to the
identity
\begin{equation}\label{wn-winf-ser}
\hat{B}^n_k( f)=\frac{1}{2^{n-1}} \sum_{m=0}^{n-1}
\binom{n-1}{m}
  \sum_{p=0}^{\infty}  \sum_{q=0}^{\infty}  (-1)^p \frac{k^{p+q}}{p! q!} B^{ m+p}_{n-1-m+q}(f).
\end{equation}
 The series on the right hand side (\ref{wn-winf-ser}) is convergent
in the natural topology mentioned at the end of Section~\ref{sec-HPWN-schm}
(for example its matrix element, in an arbitrary pair of f\/irst order number vectors,
reduces to a f\/inite sum, see~\cite{id} for more details). This gives a natural
meaning to its analytic continuation in a neighborhood of $k=0$, which is used
in the following inversion formula:
\begin{equation*}
B^n_k( f)=\sum_{\rho=0}^k  \sum_{\sigma=0}^n
\binom{k}{\rho}\binom{n}{\sigma} \frac{(-1)^{\rho}}{2^{\rho+\sigma}}
\frac{\partial^{\rho+\sigma }}{\partial {z}^{\rho+\sigma}}\Big|_{z =0}
\,\hat{B}^{k+n+1-(\rho+\sigma)}_{z}(f).
\end{equation*}

The conclusion is that, as suggested by the analogy with the classical case,
even though the two $*$-Lie algebras $w_{\infty}$ and RHPWN are dif\/ferent from
a purely algebraic point of view, their closure in a natural topology coincide.
 Moreover the explicit representations given above provide a concrete realization of the RHPWN as sesquilinear forms
on the space of the f\/irst order white noise.

The problem of realizing them as bona f\/ide closable operators on some
Hilbert space is largely open. For example in the case of the
$w_{\infty}$-operators $\hat{B}_k^n$, only for $n=2$
the Witt--Virasoro algebra, such a representation is available
(see the comment at the end of Section~\ref{gen-Fck-repr-RHPWN}).

\section{Fighting the no-go theorems}\label{pers-no-go-thms}

 A possible way out from the no-go theorems is to look for
a modif\/ication of the notion of Fock representation that keeps its main property
(algebra implies statistics) but avoids ghosts.
There is no standard rule for producing such modif\/ications: the best
one can do is to conjecture a~possible candidate through heuristic manipulations
and then try to prove that it has the desired properties.

In the following we illustrate this procedure by a two step modif\/ication of
the notion of Fock representation: the f\/irst step is not suf\/f\/icient to avoid ghosts
(this section) while the second one (following section), which improves the f\/irst
one by adding a diagonal prescription, leads to {\em bona fide} Hilbert
space representations.

We feel that the comparison between the positive and negative result
may help the reader get an intuition of ``what makes things work''.

One can show that the symbolic expressions
\begin{equation}\label{numb-gen}
B^n_k(f)=\int_{\mathbb{R}} f(t) (b_t^{\dagger})^{n-k} (b_t b_t^{\dagger})^k dt
\end{equation}
 for natural integers $n\geq k\geq0$ and functions $f$
in the test function space def\/ined by~(\ref{bdry-cond-0}), provide,
together with their adjoints which by def\/inition are denoted $B^k_n(f)$ (recall that
$n\geq k\geq0$), a~set of generators of the RHPWN algebra.

This means that, applying to these symbolic expressions the known formulae
on the combinatorics of the (Boson) creation/annihilation operators combined
with the new renormalization prescription (\ref{conv-ren}), one obtains the
RHPWN $*$-Lie algebra, that was obtained by applying the same procedure to the
symbolic expressions~(\ref{smrd-op-HP}).

Now, we want to keep track of the heuristic interpretation of $b_t b_t^{\dagger}$
as a~white noise operator. Therefore its action on the modif\/ied Fock vacuum $\Phi $
should be compatible with the usual Fock prescription $b_t\Phi =0$, i.e.\ it
should be of the form
\[
b_t b_t^{\dagger}\Phi = c_t\Phi,
\]
 where $c_t$ is a renormalization constant. Proposition~2 of \cite{[AcBou08a]}
shows that $c_t$ must be independent of $t$ so that, up to a multiplicative constant,
the action of $B^n_k(f)$ on $\Phi $ should be a multiple of $B^{n-k}_0(f)\Phi$ for
$n\geq k\geq0$ and zero for $0\leq n < k$.
This suggests the following def\/inition.

\begin{definition}\label{action}
A generalized Fock representation of the RHPWN $*$-Lie algebra
with gene\-ra\-tors~(\ref{numb-gen}) is a triple
$\{ \mathcal{H}, \Phi, \pi \}$
such that $\mathcal{H}$ is a Hilbert space, $\Phi$ a unit vector cyclic for the generators (\ref{numb-gen}), and $\pi$ is a~$*$-representation of RHPWN (from now on omitted from notations for simplicity) with the follo\-wing properties:
\begin{equation}\label{nat2}
B^n_k(f) \Phi:=
\begin{cases}
0 &\text{if $n<k$ or $n\cdot k<0$},\\
\sigma_1^kB^{n-k}_0(f) \Phi &\text{if $n>k\geq0$},\\
\displaystyle \frac{\sigma_1^k}{k+1} \int_{\mathbb{R}} f(t) dt \Phi &\text{if $n= k$},
\end{cases}
\end{equation}
 where  $\sigma_1$ is a real number depending on the renormalization.
\end{definition}

\begin{remark}
Up to a rescaling we can always assume that in (\ref{nat2})
one has $\sigma_1=1 $ so that, for $n,k\in\mathbb{N}$ and test functions $f$:
\begin{gather}\label{Fk-pres-n1a}
B^n_k(f) \Phi:= 0  \qquad  \text{if \ \ $n<k$},
\\
\label{Fk-pres-n1b}
B^n_k(f) \Phi:=\frac{1}{n+1} \int_{\mathbb{R}} f(t) dt \Phi \qquad  \text{if \  \ $n= k$},
\\
\label{Fk-pres-n1c}
B^n_k(f) \Phi:= B^{n-k}_0(f) \Phi \qquad \text{if \ \ $n>k\geq0$}.
\end{gather}
\end{remark}

In the following we will assume that the test function $\chi_I$,
in (\ref{charfunc}), is such that $I\subseteq \mathbb{R}\setminus \{0\}$ is an interval and $\chi_I(x)=1$ if $x\in I$, $\chi_I(x)=0$ if $x\notin I$. We also suppose that the Lebesgue measure of $I$ is suf\/f\/iciently large so that the no-go theorems do not apply. Under these assumptions the $*$-Lie algebra generated by the $B^n_k$ and $B^k_n$ can be considered as a ``one-mode'' realization of the RHPWN  $
*$-Lie algebra.

\begin{definition}\label{L}\qquad
\begin{enumerate}\itemsep=0pt
\item[$(i)$]  $\mathcal{L}_1$ is the $*$-Lie algebra generated by $B^1_0$ and $B^0_1$, i.e., $\mathcal{L}_1$ is the linear span of $\{B^1_0,B^0_1,B^0_0\}$ (the usual oscillator algebra).

\item[$(ii)$] $\mathcal{L}_2$ is the $*$-Lie algebra generated by $B^2_0$ and $B^0_2$, i.e., $\mathcal{L}_2$ is the linear span of $\{B^2_0,B^0_2,B^1_1\}$ (the usual quadratic algebra, isomorphic to $sl(2,\mathbb{R})$).

\item[$(iii)$] For $n\in\{3,4,\dots\}$, $\mathcal{L}_n$ is the $*$-Lie subalgebra of
RHPWN generated by $B^n_0$ and $B^0_n$. It is the linear span of the operators
of the form $B^x_y$ where $x-y=k n$,  $k\in\mathbb{Z}\setminus\{0\}$,
and of the number operators $B^{x}_{x}$ with $x\geq n-1$.
\end{enumerate}
\end{definition}

 The proof of the following theorem can be found in~\cite{[AcBo06]}.

\begin{theorem}\label{NoGo}
Let $n\geq3$ and suppose that a generalized Fock representation $\{\mathcal{F}_n, \Phi \}$, of $\mathcal{L}_n$, in the sense of Definition~{\rm \ref{action}}, exists. Then it contains both $B^{n}_0 \Phi$ and $B^{2n}_0 \Phi$. In particular, if the test function space includes functions whose support has arbitrarily small Lebesgue measure, then $\mathcal{L}_n$ does not admit a generalized Fock representation in the sense of  Definition~{\rm \ref{action}}.
\end{theorem}

\section{Further generalizations of the Fock representation\\ for the RHPWN algebra}\label{gen-Fck-repr-RHPWN}

 In this section we prove that a further strengthening of the notion of Fock representation for the RHPWN algebra leads to well def\/ined unitary representations.
The idea of the construction is the following.

\begin{definition}
A generalized Fock representation of the RHPWN $*$-Lie algebra with
gene\-ra\-tors~(\ref{numb-gen}) is def\/ined, as in Def\/inition~\ref{action},
to be a triple $\{ \mathcal{H}, \Phi, \pi \}$ satisfying the two conditions~(\ref{Fk-pres-n1a}) and (\ref{Fk-pres-n1b})
 and replacing (\ref{Fk-pres-n1c}) by (in the notation (\ref{charfunc})):
\begin{equation}\label{gen-Fk}
 B^{n+x}_{x} (B^n_0)^{N}\Phi = B^{n}_{0} (B^n_0)^{N}\Phi
, \qquad \forall \, n,x,N\in \mathbb{N}.
\end{equation}
\end{definition}

One easily verif\/ies that condition (\ref{gen-Fk}) implies that:
\[
B^{n-1}_{n-1}(B^n_0)^{k}\Phi = \left({\frac{\mu(I)}{ n}} + k n(n-1)\right) (B^n_0)^{k}\Phi
\]
or, writing explicitly the test function, $\forall\,k,n\in\Bbb N$:
\begin{equation*}
B^{n-1}_{n-1}(\chi_{I}) (B^n_0(\chi_{I}))^k \Phi:=
\left(\frac{\mu(I)}{n}+k n (n-1)\right) (B^n_0(\chi_{I}))^k \Phi.
\end{equation*}

The prescription that the vectors $B^n_0(\chi_I)^k\Phi$ are total in ${\cal H}$ uniquely determines this representation up to isomorphism. In fact these prescriptions
uniquely f\/ix the inner product among the higher order particle vectors to be given by:
\begin{equation*}
\langle (B^n_0(\chi_{I}))^k \Phi, (B^n_0(\chi_{I}))^m \Phi \rangle=
\delta_{m,k} k! n^k \prod_{i=0}^{k-1}\left(\mu(I)+\frac{n^2 (n-1)}{2} i\right)
\end{equation*}
with
\begin{equation*}
\langle (B^n_0(\chi_{I}))^k \Phi, (B^n_0(\chi_{J}))^m \Phi \rangle=0
\end{equation*}
if $I$ and $J$ are disjoint.

 Def\/ining the $n$-th order exponential vectors
by analogy with the f\/irst and the second order case, i.e.
\begin{equation*}
\psi_n(\phi):=\prod_{i=1}^m e^{b_i B^n_0(\chi_{I_i})} \Phi,
\end{equation*}
where $\phi$ is the compact support step function
\begin{equation*}
\phi:=\sum_{i=1}^m b_i \chi_{I_i}
\end{equation*}
for $n=1$ one f\/inds
\begin{equation}\label{n=1}
\langle \psi_1(f),\psi_1(g)\rangle_1:=
e^{\int_{\mathbb{R}^d}  \bar f(t) g(t) dt }=e^{\langle f,g\rangle_{L^2(\Bbb R^d)}}
\end{equation}
and for $n\geq2$
\begin{equation}\label{n>1}
\langle \psi_n(f),\psi_n(g)\rangle:=e^{ -\frac{2}{n^2 (n-1)}
\int_{\mathbb{R}}  \ln\left(1-\frac{n^3 (n-1)}{2}   \bar f(t) g(t)\right) dt},
\end{equation}
 where the integral in (\ref{n>1}) exists under the condition
\[
\sup_{t\in\Bbb R^d}|f(t)|< \frac{1}{n} \sqrt{\frac{2}{n (n-1)}},\qquad
\sup_{t\in\Bbb R^d}|g(t)|<\frac{1}{n} \sqrt{\frac{2}{n (n-1)}}.
\]
The positivity of the scalar product is now clear because
(\ref{n=1}) is the formula for the inner product of two exponential vectors in the usual (f\/irst order) Fock space while (\ref{n>1}) coincides (up to rescalings)
with the formula, found in \cite{[AcSk99b]}, for the inner product of two exponential
vectors in the quadratic case.

The action of the RHPWN operators on the exponential vectors is given by:
\begin{gather*}
B_0^n(f) \psi_n(g)=\frac{\partial}{\partial \epsilon}\Big|_{\epsilon=0} \psi_n(g+\epsilon f),
\\
B^0_n(f) \psi_n(g)=n \int_{\mathbb{R}} f(t) g(t) dt\, \psi_n(g)
+\frac{n^3 (n-1)}{2} \frac{\partial}{\partial \epsilon}\Big|_{\epsilon=0}
\psi_n(g+\epsilon f g^2)
\end{gather*}
for all test functions $f:=\sum_i a_i \chi_{I_i}$ and $g:=\sum_i b_i \chi_{I_i}$
with  $ I_i \cap I_j= \oslash $ for $i\neq j$, and for all $n\geq1$.

The emergence of the quadratic Fock space in this context can be explained, a posteriori,
as follows.  Denote
\[
c:= {\frac{\mu(I)}{ n}}, \qquad  q_k := k n(n-1)
, \qquad  (B^n_0)^k\Phi =: |k\rangle,
\]
 so that
\begin{gather*}
B^n_0|k\rangle=|k+1\rangle,
\\
B^{n-1}_{n-1}|k\rangle = B^{n-1}_{n-1}(B^n_0)^k\Phi=(c + q_k)(B^n_0)^k\Phi
=(c + q_k)|k\rangle,
\end{gather*}
i.e.\ the restriction of $B^{n-1}_{n-1}$ on the algebraic linear span ${\cal F}_n$
of the vectors $(|k\rangle)$, which are clearly mutually orthogonal, takes the form
\[
B^{n-1}_{n-1}=\sum (c + q_k)|k\rangle\langle k| = c\cdot 1 + q_N,
\]
where
\[
q_N |k\rangle := q_k |k\rangle
\]
 Consequently
\begin{gather*}
[B^{n-1}_{n-1},B^n_0]|k\rangle=B^{n-1}_{n-1}B^n_0|k\rangle-B^n_0B^{n-1}_{n-1}|k\rangle
=B^{n-1}_{n-1}|k+1\rangle-q_kB^n_0|k\rangle
\\
\phantom{[B^{n-1}_{n-1},B^n_0]|k\rangle}{}
=(q_{k+1}-q_k)|k+1\rangle   = n(n-1)B^n_0|k\rangle.
\end{gather*}
In conclusion, on  ${\cal F}_n$ the following commutation relations hold:
\begin{gather*}
[B^0_n,B^n_0]=n^2B^{n-1}_{n-1}=n^2c1+n^2q_N,
\\
[B^{n-1}_{n-1},B^n_0]=n(n-1)B^n_0
\end{gather*}
and, up to rescalings, these are the relations def\/ining a central extension of
$sl(2,\mathbb{R})$, which is precisely the one-particle algebra of the quadratic white noise.

The construction given in \cite{[AcBou08a]} of the generalized Fock representation (in the sense specif\/ied above) and the determination of the corresponding statistics was based on algebraic techniques.

An analytical construction of the corresponding generalized Fock space, within an extension of Hida theory of white noise calculus to the negative binomial process, has recently been obtained by Barhoumi, Ouerdiane and Rihai \cite{BarOuerRia08}. This analytical
construction is particularly interesting because it shows that the reduction with the quadratic case does not destroy the connection with the Virasoro algebra. In fact these authors prove that, for $n=2$, the symbols of the operators $\hat{B}^n_k( f)$, def\/ined by~(\ref{wn-winf-ser}) (i.e.\ their matrix elements in the higher order exponential vectors)
ef\/fectively def\/ine closable operators in the quadratic Fock space while this is not true
for $n>2$.

\subsection{Classical stochastic processes associated with the representation}\label{Cl-stoch-pr-gen-Fk-rep-RHPWN}

In this section we work in dimension $1$ (i.e.\ $d=1$).
For $n\geq 1$ consider the operator process:
\begin{equation}\label{field-proc1}
\{ B^n_0(\chi_{[0,t]}) + B^0_n(\chi_{[0,t]})   : \ t\in \mathbb{R}_+ \}.
\end{equation}
  It is not dif\/f\/icult to verify that the family (\ref{field-proc1})
is commutative. Therefore its vacuum distributions def\/ine,
as described in Section~\ref{Prob-impl}, a classical stochastic process.
The following theorem identif\/ies these processes as continuous binomial
(or Beta) processes. This is a subclass of the Meixner processes which,
in their turn, are a special family of stationary independent increment (or L\'evy)
processes.

\begin{theorem}\label{mgf} The vacuum moment generating functions
of the operator process \eqref{field-proc1}  is given, for $n=1$, by:
\begin{equation*}
\langle  \Phi, e^{s\,(B^1_0(t)+B^0_1(t))} \Phi \rangle_1=e^{\frac{s^2}{2} t}
,\qquad    s\in [0,\infty),
\end{equation*}
i.e.\  the process  $\{B^1_0(t)+B^0_1(t)  : \ t\geq0\}$ is
the standard classical Brownian motion. While for $n\geq2$ one f\/inds:
\begin{equation*}
\langle \Phi,  e^{s (B^n_0(t)+B^0_n(t))} \Phi \rangle_n=
\left(\sec \left(\sqrt{\frac{n^3 (n-1)}{2}} s\right)\right)^{ \frac{2 n t}{n^3 (n-1)}}
,\qquad  s\in [0,\infty),
\end{equation*}
  i.e.\  $\{B^n_0(t)+B^0_n(t)   : \ t\geq0\}$ is for each $n$
a continuous binomial (or Beta) process with density
\begin{equation*}
\mu_{t, n}(x)=
p_{\frac{2 n t}{n^3 (n-1)}   }(x)=
\frac{2^{\frac{2 n t}{n^3 (n-1)}  -1}}{2 \pi}
B\left(\frac{\frac{2 n t}{n^3 (n-1)}  +i x}{2},
\frac{\frac{2 n t}{n^3 (n-1)}  -i x}{2}\right),
\end{equation*}
 where $B(a,c)$ is the Beta function with parameters $a$, $c$:
\begin{equation*}
B(a,c)=\frac{\Gamma(a) \Gamma(c)}{\Gamma(a+c)}=\int_0^1 x^{a-1} (1-x)^{c-1} dx
,\qquad \Re a>0,\qquad \Re c>0.
\end{equation*}
\end{theorem}

 Notice incidentally that the formula postulated by Veneziano for
the scattering amplitude of some strong interactions was also def\/ined in terms
of the Euler beta function (see \cite[p.~373]{[GrSchwWi87-I]}).

\section{Central extensions and renormalization}\label{ce}

 The analogy with the Virasoro algebra naturally suggests
 the investigation of the existence of Hilbert space representations
not of the RHPWN $*$-Lie algebra itself but of some of its central extensions.

But the RHPWN $*$-Lie algebra is a second quantized and renormalized version of the full oscillator algebra (FOA).
Thus a natural preliminary problem is to construct central extensions
of the FOA.

Since, in its turn, the FOA is the universal enveloping algebra of
the Heisenberg algebra and since a central extension of an algebra automatically provides an extension (even if usually not central)
of its universal enveloping algebra, this leads to the further
preliminary problem to construct central extensions of
the Heisenberg algebra.

 In the paper \cite{AccBouCEHeis} the following results were established:
\begin{enumerate}\itemsep=0pt

\item[$(i)$] There exists a one (complex) parameter family of nontrivial central extensions of the Heisenberg $*$-Lie algebra.

\item[$(ii)$] These nontrivial central extensions of the Heisenberg $*$-Lie algebra have a boson realization within the Schr\"odinger algebra.

\item[$(iii)$] This boson realization can be used to compute the vacuum characteristic function of the f\/ield operators in the Fock representation of the Schr\"odinger algebra.

\item[$(iv)$] The above mentioned boson realization can also be used to obtain a second quantized version of a quadratic boson algebra
which cannot be deduced from the constructions in~\cite{[AcLuVo99]} and~\cite{[AcFrSk00]}. This consequence is quite nontrivial due to
the no-go theorems: it conf\/irms the strict connection between
central extensions and renormalization even if the deep roots
of this connection have yet to be clarif\/ied.
In the following we brief\/ly recall the central extensions mentioned
in item $(i)$ above.
\end{enumerate}

  Recall (cf.~\cite{Fuchs,9+}) that, if $L$  and $\widetilde L$ are two
complex Lie algebras, $\widetilde L$ is called a one-dimensional
\textit{central extension} of $L$ with \textit{central element} $E$ if
\[
\lbrack l_1, l_2 \rbrack_{\widetilde L}=
\lbrack l_1, l_2 \rbrack_L +\phi(l_1,l_2) E,\qquad
\lbrack l_1,E \rbrack_{\widetilde L}=0
, \qquad \forall \, l_1, l_2 \in L,
\]
  where $\lbrack \cdot, \cdot \rbrack_{\widetilde L}$ and
$\lbrack \cdot, \cdot\rbrack_L$ are the Lie brackets in  $\widetilde L$ and $L$ respectively and $\phi:L\times L\mapsto \mathbb{C}$ is a~$2$-cocycle on $L$, i.e.\ a bilinear form
satisfying the skew-symmetry condition
\[
\phi (l_1,l_2)=-\phi(l_2,l_1)
\]
 and the Jacobi identity
\[
\phi(\lbrack l_1,l_2 \rbrack_{ L},l_3)+
\phi(\lbrack l_2,l_3 \rbrack_{ L},l_1)+
\phi(\lbrack l_3,l_1 \rbrack_{ L},l_2)=0.
\]
 A  central extension is \textit{trivial} if there exists
a linear function $f: L \mapsto \mathbb{C}$ satisfying for all $l_1, l_2 \in L$
\[
\phi (l_1,l_2)=f(\lbrack l_1,l_2 \rbrack_{L}).
\]

\subsection{Central extensions of the Heisenberg algebra}\label{sec-Centr-ext-Heis-alg}

 While all central extensions of the Oscillator algebra
(i.e.\ the $*$-Lie subalgebra of the full oscillator algebra
generated by $B^1_0$, $B^0_1$, $B^0_0$, and $B^1_1$)
as well as those of the Square of White Noise algebra
(generated by $B^2_0$, $B^0_2$, and $B^1_1$ and isomorphic
to $sl(2,\mathbb{R})$) are trivial, this is not true for the Heisenberg algebra, i.e.\ the $*$-Lie subalgebra of the full oscillator algebra generated by $B^1_0$, $B^0_1$ and $B^0_0$.
In fact all central extensions of the Heisenberg algebra are described as follows:
\begin{gather*}
\left(B_0^1\right)^{*} = B_1^0,\qquad \left(B_1^0\right)^{*} = B_0^1, \qquad \left(B_0^0\right)^{*} = B_0^0,
\\
\lbrack B^0_1,B^1_0\rbrack=
B^0_0 + \lambda  E, \qquad \lbrack B^0_0,B^1_0\rbrack=  z  E,\qquad \lbrack B^0_1,B^0_0\rbrack= \bar z  E,
\end{gather*}
where $z\in\mathbb{C}$ and $\lambda  \in \mathbb{R}$ are arbitrary constants. A central extension of the Heisenberg  $*$-Lie algebra is trivial if and only if $z=0$.

Up to algebraic (but not stochastic) isomorphism there is one nontrivial central
extension which belongs to the list of $15$ real $4$-dimensional Lie algebras
in the classif\/ication due to Kruchkovich~\cite{[Kruchk54],Ovando}. One boson realization
of this algebra (see~\cite{FS96}) is generated by $\{ q^2,q,p,1\}$. This shows that
it can be identif\/ied to a subalgebra of the Schr\"odinger algebra (for whose
current algebra over $\mathbb{R}$ we know that a no-go theorem holds). The presence
of $q^2$ shows that a~renorma\-li\-zation is required to give a meaning to the
associated current algebra. However the simultaneous presence of $q$ and $p$ shows that,
if a Fock representation of this algebra exists, then it cannot be realized in the RSWN
$*$-Lie algebra discussed in Section~\ref{sec-quadr-pow}. However from the end of Section~\ref{sec-Bos-alg} we know that the associated $*$-Lie algebra is well def\/ined and
one can show that the general construction of \cite{Guichardet} and  \cite{Partha-Schmidt}
can be applied, hence $*$-representations can be constructed. It is however not known if any
of these representations can be identif\/ied in a~natural way to the Fock representation
of this algebra (which, if existing as a $*$-representation, is uniquely determined up
to unitary isomorphism).

\subsection[Central extensions of the RHPWN and of the $w_{\infty}$ algebra]{Central extensions of the RHPWN and of the $\boldsymbol{w_{\infty}}$ algebra}

  The $*$-algebras RHPWN and $w_{\infty}$ are too large to admit nontrivial central extensions. In \cite{AccBouCE} the following results were proved:
\begin{enumerate}\itemsep=0pt
\item[$(i)$] All central extensions of the RHPWN $*$-Lie algebra are trivial.

\item[$(ii)$] All central extensions of the higher order $w_{\infty}$ $*$-Lie algebra (i.e.\ with the Witt--Virasoro sector removed) are trivial.
\end{enumerate}

 The statement $(i)$ is new, the statement $(ii)$ provides a new proof of a result due to Bakas \cite{6} who proved that the coef\/f\/icients of the central terms of a suitable contraction of the Zamolodchikov Lie algebra $W_N$ go to zero (cf.~\cite{9,[Zamol85]}) as $N\rightarrow\infty$.
From this he could conclude that all cocycles, which arise from that
f\/inite dimensional approximation, are trivial. Our proof, being based on a purely algebraic analysis of the $2$-cocycles of $w_{\infty}$, is more general and direct.

One might hope that the $*$-Lie algebras $\mathcal{L}_n$, of the RHPWN $*$-Lie algebra, introduced in Section~\ref{pers-no-go-thms} above, admit nontrivial central extensions.
This seems to be unlikely due to the recent discovery (see~\cite{AccBouCE}) that, in the
family of natural subalgebras of $w_{\infty}$, only the Virasoro algebra admits
nontrivial central extensions. A similar result for subalgebras of RHPWN is not available
at the moment, but we hope to come back to this point soon.

\section{Conclusions}

The identif\/ication of the (closures of) the RHPWN and the $w_{\infty}$-$*$-Lie algebras
suggests that these algebras have a canonical mathematical meaning.
However our program of identifying the elements of these algebras to renormalized powers of white noise can be considered realized only in the quadratic case.
An important guiding principle that we have learned from this case is that,
if this program can be realized, then in the representation space there should be
a unit vector $\Phi$ such that the $\Phi$-moments of the classical process given by
the renormalized $n$-th power of $(b^+_t+b_t)$ should be an independent increment process
whose distribution is the $n$-th power of the standard Gaussian.
This brings a connection with an old open problem of classical probability.
In fact, while it is known that even powers of the standard Gaussian are inf\/initely divisible,
the same statement for odd powers ($\geq 3$) is not known and experts conjecture that
the answer is negative.
This suggests that one could use the no-go theorems to deduce a negative answer to this classical conjecture based on quantum probabilistic techniques.
Moreover fortunately, both for RHPWN and for $w_{\infty}$, the even powers form
a $*$-Lie subalgebra, so that one can restrict one's attention to even powers.
This gives the advantage that one knows a priori a natural candidate for
the representation space: i.e.\ for each $n\in\mathbb N$, the $L^2$-space of the
independent increment stationary process corresponding to the $2n$-th power of the
standard Gaussian.

Our strategy to take advantage of this information consists in the complete inversion
of the strategy pursued from 1999 up to now, namely: up to now we have pursued one
of the basic tenets of quantum probability, {\it algebra implies statistics},
in the sense that we have tried to guess a reasonable def\/inition of {\it Fock}
(lowest weight) representation and to deduce the statistics from it on the lines
outlined in Section~\ref{sec-Fk-rep} and after equation (\ref{df-quad-Fk-st-A}).

From now on we will pursue the other basic tenet of quantum probability,
{\it statistics implies algebra}, and starting from the $L^2$-space of the
independent increment stationary process corresponding to the $2n$-th power of the
standard Gaussian we will apply the theory of interacting Fock spaces to deduce the
quantum decomposition of this classical process and the commutation relations
canonically associated to the principal Jacobi sequence of this distribution (which is
symmetric so that the secondary Jacobi sequence vanishes identically).

In this direction we will surely benef\/it from the results developed by Y.~Berezansky
and his school on the extension of classical white noise theory to a general class
of L\'evy processes (Jacobi f\/ields, see \cite{[BereLytvMier03],[Lytv02]}).
The $L^2$-spaces of these processes are naturally isomorphic to a~class of $1$-mode
type interacting Fock spaces which includes the even powers of the Gaussian.
Although  naturally isomorphic the two realizations are dif\/ferent and the interacting
Fock space one is simpler to handle from the algebraic point of view.

Another direction might be to look for representations dif\/ferent from the Fock one (cf.~\cite{Kac, [Kac78],[Kac79]}).
This is surely a direction worth investigating. However at the moment our knowledge of
such representations is rather limited even in the quadratic case. In fact, as already
mentioned in Section~\ref{sec-Fk-rep}, even in the f\/irst order case our explicit
control of these representations is restricted to the Gaussian (or quasi-free) ones.

As often in mathematics what has been understood is a tiny fraction of what
one would like to understand. However the landscape that has emerged along this path
is so intriguing and promising that it constitutes a stimulus to meet this challenge.

\subsection*{Acknowledgements}

We want to express our gratitude to the f\/ive referees who examined our paper. Their comments clearly indicate that they have gone through it carefully and the present version of the paper has greatly benef\/itted from them.

\pdfbookmark[1]{References}{ref}
\LastPageEnding


\begin{thebibliography}{99}

\footnotesize\itemsep=0pt


\bibitem{[AcAmFr02]}
Accardi L., Amosov G.,  Franz U.,
Second quantized automorphisms of the renormalized square of white noise (RSWN) algebra
{\it Infin. Dimens. Anal. Quantum Probab. Relat. Top.} {\bf 7} (2004), 183--194.

\bibitem{[AcArVo03]}
Accardi L., Arefeva I., Volovich I.V.,
Quadratic Fermi f\/ields, unpublished manuscript, 2003.


\bibitem{AccBouCEHeis}
Accardi L., Boukas A.,
Central extensions of the Heisenberg algebra,
{\it Infin. Dimens. Anal. Quantum Probab. Relat. Top.}, submitted,
\href{http://arxiv.org/abs/0810.3365}{arXiv:0810.3365}.


\bibitem{AccBouCE}
Accardi L., Boukas A.,
Central extensions of white noise $*$-Lie algebras,
 {\it Infin. Dimens. Anal. Quantum Probab. Relat. Top.}, submitted.


\bibitem{[AcBou08a]}
Accardi  L., Boukas  A.,
Fock representation of the renormalized higher powers of white noise and the centerless Virasoro (or Witt)--Zamolodchikov-$w_{\infty}$ $*$-Lie algebra,
{\it J. Phys. A: Math. Theor.} {\bf 41} (2008), 304001, 12~pages, \href{http://arxiv.org/abs/0706.3397}{arXiv:0706.3397}.

\bibitem{[AcBou05a]}
Accardi L., Boukas A.,
Higher powers of $q$-deformed white noise,
{\it Methods Funct. Anal. Topology} {\bf 12} (2006), 205--219.

\bibitem{[AcBou05g]}
Accardi L.,  Boukas A.,
It\^o calculus and quantum white noise calculus,
Proceedings of ``The 2005 Abel Symposium, Stochastic Analysis and
Applications'' (Symposium in Honor of Kiyosi It\^o,
on the occasion of his 90th birthday) (July 29 -- August 4, 2005, Oslo,
     Norway),
Editors  F.E.~Benth, G.~Di Nunno, T.~Lindstrom, B.~Oksendal and T.~Zhang,
Springer, Berlin, 2007   Vol.~2, 7--51.


\bibitem{[AcBo07]}
Accardi  L., Boukas  A.,
Lie algebras associated with the renormalized higher powers of white noise,
{\it Commun. Stoch. Anal.}  {\bf 1} (2007), 57--69.


\bibitem{[Ac00c]}
Accardi L.,
Quantum probability: an introduction to some basic ideas and trends,
in Stochastic models, II   (Guanajuato, 2000),  Editors  D.~Hernandez, J.A.~Lopez-Mimbela and R.~Quezada,
{\it Aportaciones Mat. Investig.}, Vol.~16, Soc. Mat. Mexicana, Mexico, 2001, 1--128.

\bibitem{[AcBou06a]}
Accardi L., Boukas A.,
Recent advances in quantum white noise calculus,
in Quantum Information and Computing,
{\it Quantum Probability and White Noise Analysis}, Vol.~19, World Scientif\/ic, 2006, 18--27.

\bibitem{id}
Accardi L., Boukas A.,
Renormalized higher powers of white noise and the Virasoro--Zamolodchikov-$w_{\infty}$ algebra,
{\it Rep. Math. Phys.} {\bf 61} (2008), 1--11,
\href{http://arxiv.org/abs/hep-th/0610302}{hep-th/0610302}.

\bibitem{[AcBou06]}
Accardi  L., Boukas  A.,
Renormalized higher powers of white noise (RHPWN) and conformal f\/ield theory,
{\it Infin. Dimens. Anal. Quantum Probab. Relat. Top.} {\bf 9} (2006),  353--360,
\href{http://arxiv.org/abs/math-ph/0608047}{math-ph/0608047}.


\bibitem{[AcBou01e]}
Accardi L., Boukas A.,
Square of white noise unitary evolutions on boson Fock space,
in Proceedings of the International Conference on Stochastic Analysis and Applications in Honor of Paul Kree (October 22--27, 2001, Hammamet, Tunisie), Editors S.~Albeverio, A.~Boutet de Monvel and H.~Ouerdiane, Kluwer Acad. Publ., Dordrecht, 2004, 267--301.

\bibitem{[AcBo06]}
Accardi  L., Boukas  A.,
The emergence of the Virasoro and $w_{\infty}$ Lie algebras through the renormalized higher powers of quantum white noise,
{\it Int. J. Math. Comput. Sci.} {\bf 1} (2006), 315--342,
\href{http://arxiv.org/abs/math-ph/0607062}{math-ph/0607062}.

\bibitem{[AcBou01d]}
Accardi L., Boukas A.,
The semi-martingale property of the square of white noise integrators,
in Stochastic Partial Dif\/ferential Equations and Applications (Trento, 2002),  Editors G.~Da Prato and L.~Tubaro,
{\it Lecture Notes in Pure and Appl. Math.}, Vol.~227, Dekker, New York, 2002, 1--19.

\bibitem{[AcBou03a]}
Accardi L., Boukas A.,
Unitarity conditions for the renormalized square of white noise,
{\it Infin. Dimens. Anal. Quantum Probab. Relat. Top.} {\bf 6} (2003), 197--222.

\bibitem{[AcBou04c]}
Accardi L., Boukas A.,
White noise calculus and stochastic calculus,
in Proceedings of the International Conference ``Stochastic Analysis: Classical and Quantum,
Perspectives of White Noise Theory'' (November 1--5, 2003, Meijo University, Nagoya),
Editors T.~Hida and K.~Saito, World Sci. Publ., Hackensack, NJ, 2005, 260--300.

\bibitem{[AcBouFr06]}
Accardi L., Boukas A., Franz U.,
Renormalized powers of quantum white noise,
{\it Infin. Dimens. Anal. Quantum Probab. Relat. Top.} {\bf 9} (2006), 129--147.

\bibitem{[AcBou01f]}
Accardi L., Boukas A., Kuo H.-H.,
On the unitarity of stochastic evolutions driven by the square of white noise,
{\it Infin. Dimens. Anal. Quantum Probab. Relat. Top.} {\bf 4} (2001), 579--588.


\bibitem{[AcFrSk00]}
Accardi L., Franz U., Skeide M.,
Renormalized squares of white noise and other non-Gaussian noises as L\'evy processes on real Lie algebras,
{\it Comm. Math. Phys.} {\bf 228} (2002), 123--150.

\bibitem{[AcHiKu01]}
Accardi L., Hida T., Kuo H.H.,
The It\^o table of the square of white noise,
{\it Infin. Dimens. Anal. Quantum Probab. Relat. Top.} {\bf 4} (2001), 267--275.

\bibitem{[AcLuOb96]}
Accardi L., Lu Y.G., Obata N.,
Towards a non-linear extension of stochastic calculus,
 in  Quantum Stochastic Analysis and Related Fields (Kyoto, 1995),
  Editors N.~Obata,
 {\it S\={u}rikaisekikenky\={u}sho K\={o}ky\={u}roku}, no.~957, Research Institute for Mathematical
Sciences, Kyoto, 1996, 1--15.


\bibitem{[AcLuVo95b]}
Accardi L.,  Lu  Y.G.,  Volovich I.,
Non-linear extensions of classical and quantum stochastic calculus
and essentially inf\/inite-dimensional analysis,
in Probability Towards 2000 (October 2--6, 1995, Columbia University, New York),
Editors L.~Accardi and C. Heyde,
{\it Lecture Notes in Statist.}, Vol.~128, Springer, New York, 1998, 1--33.


\bibitem{[AcLuVo02]}
Accardi L.,  Lu  Y.G.,  Volovich I.,
Quantum theory and its stochastic limit, Springer-Verlag, Berlin, 2002.

\bibitem{[AcLuVo99]}
Accardi L.,  Lu  Y.G.,  Volovich I.,
White noise approach to classical and quantum stochastic calculi,
Lecture Notes of the Volterra International School, Trento, Italy, 1999.

\bibitem{[AcPeRo04]}
Accardi L., Pechen A.,  Roschin R.,
Quadratic KMS states on the RSWN algebra,  unpublished manuscript, 2004.


\bibitem{[AcRo05]}
Accardi L., Roschin R.,
Renormalized squares of Boson f\/ields,
{\it Infin. Dimens. Anal. Quantum Probab. Relat. Top.} {\bf 8} (2005), 307--326.

\bibitem{[AcSk99b]}
Accardi L., Skeide M.,
On the relation of the square of white noise and the f\/inite dif\/ference algebra,
{\it Infin. Dimens. Anal. Quantum Probab. Relat. Top.} {\bf 3} (2000), 185--189.


\bibitem{[AcVo97]}
Accardi L., Volovich I.V.,
Quantum white noise with singular non-linear interaction,
in Developments of Inf\/inite-Dimensional Noncommutative Analysis (Kyoto, 1998), Editor N.~Obata,
{\it S\={u}rikaisekikenky\={u}sho K\={o}ky\={u}roku}, no.~1099,  Research Institute for
Mathematical Sciences, Kyoto, 1999, 61--69.


\bibitem{[Arv03]}
Arveson W., Noncommutative dynamics and $E$-semigroups, {\it  Springer Monographs in Mathematics}, Springer-Verlag, New York, 2003.

\bibitem{[Ayed06]}
Ayed W.,
White noise approach to quantum stochastic calculus,
Thesis Doctor of Mathematics, University of Tunis El Manar, Tunis, 2006.

\bibitem{6}
Bakas I.,
The structure of the $W_{\infty}$ algebra,
{\it  Comm. Math. Phys.} {\bf 134} (1990), 487--508.

\bibitem{BK91}
Bakas I., Kiritsis  E.B.,
Structure and representations of the $W_{\infty}$ algebra,
{\it  Progr. Theoret. Phys. Suppl.} (1991),  no.~102, 15--37.

\bibitem{BarOuerRia08}
 Barhoumi A.,  Ouerdiane H.,  Riahi H.,
Representations of the Witt and square white noise algebras through the renormalized powers of negative binomial quantum white noise,
{\it Infin. Dimens. Anal. Quantum Probab. Relat. Top.} {\bf 11} (2008), 223--250.

\bibitem{[BereLytvMier03]}
Berezansky Yu.M., Lytvynov E., Mierzejewski D.A.,
The Jacobi f\/ield of a L\'evy process,
{\it  Ukrain. Mat. Zh.} {\bf 55} (2003), 706--710 (English transl.: {\it Ukrainian Math. J.} {\bf 55} (2003), 853--858).

\bibitem{[BogSh80]}
Bogoliubov N.N., Shirkov D.V.,
Introduction to the theory of quantized f\/ields, John Wiley \& Sons, 1980.

\bibitem{Bou91}
Boukas A.,
An example of a quantum exponential process,
{\it Monatsh. Math.} {\bf 112} (1991), 209--215.


\bibitem{4a}
El Kinani E.H., Akhoumach K., Generalized Clif\/ford algebras and certain inf\/inite dimensional Lie algebras,
{\it Adv. Appl. Clifford Algebras} {\bf 10} (2000), 1--6.

\bibitem{pfein}
Feinsilver P.,
Discrete analogues of the Heisenberg--Weyl algebra,
{\it Monatsh. Math.} {\bf 104} (1987), 89--108.

\bibitem{Fein}
Feinsilver P.J.,  Schott R.,
 Algebraic structures and operator calculus, Vols. I and III, Kluwer Academic Publishers Group, Dordrecht, 1993.

\bibitem{FS96}
Feinsilver P.J.,  Schott R.,
Dif\/ferential relations and recurrence formulas for representations of Lie groups,
{\it Stud. Appl. Math.} {\bf 96} (1996), 387--406.


\bibitem{Fuchs}
Fuchs J., Schweigert C.,
Symmetries, Lie algebras and representations. A graduate course for physicists, {\it Cambridge Monographs on Mathematical Physics}, Cambridge University Press, Cambridge, 1997.



\bibitem{[GrSchwWi87-I]}
Green M.B., Schwarz J.H., Witten E.,
Superstring theory, Vol.~I, Cambridge University Press, Cambridge, 1988.

\bibitem{Guichardet}
Guichardet A.,
Symmetric Hilbert spaces and related topics. Inf\/initely divisible positive def\/inite functions. Continuous products and tensor products. Gaussian and Poissonian stochastic processes, {\it  Lecture Notes in Mathematics}, Vol.~261, Springer-Verlag, Berlin~-- New York, 1972.


\bibitem{[Hida92]}
Hida T.,
Selected papers of Takeyuki Hida, Editors L.~Accardi, H.H.~Kuo, N.~Obata, K.~Sait\^o, Si~Si and L.~Streit, World Scientif\/ic Publishing Co., Inc., River Edge, NJ, 2001.

\bibitem{[Ivanov79]}
Ivanov V.K.,
The algebra of elementary generalized functions,
{\it Dokl. Akad. Nauk SSSR} {\bf  246} (1979), 805--808 (English transl.: {\it Soviet Math. Dokl.} {\bf  20} (1979), 553--556).


\bibitem{Kac}
Kac V.G.,  Raina A.K.,
Bombay lectures on highest weight representations of inf\/inite dimensional Lie algebras, {\it Advanced Series in Mathematical Physics}, Vol.~2, World Scientif\/ic Publishing Co., Inc., Teaneck, NJ, 1987.

\bibitem{[Kac78]}
Kac V.G.,
Highest weight
    representations   of inf\/inite-dimensional Lie algebras,
in Proceedings of the International Congress of Mathematicians (Helsinki, 1978), Acad. Sci. Fennica, Helsinki, 1980, 299--304.

\bibitem{[Kac79]}
Kac V.G., Contravariant form for
   inf\/inite-dimensional Lie algebras and superalgebras,
in Proceedings of the Seventh International Colloquium and Integrative Conference on Group Theory and Mathematical Physics (September 11--16, 1978, University of Texas, Austin),
Editors  W.~Beigelb\"ock,  A.~B\"ohm and E.~Takasugi,
 {\it Springer Lecture Notes in Physics}, Vol.~94, Springer-Verlag, Berlin~-- New York, 1979, 441--445.


\bibitem{8}
Ketov S.V.,
Conformal f\/ield theory, World Scientif\/ic Publishing Co., Inc., River Edge, NJ, 1995.

\bibitem{[Kruchk54]}
Kruchkovich G.I.,
Classif\/ication of three-dimensional Riemannian spaces according
to groups of motions, {\it Uspehi Matem. Nauk} {\bf 9} (1954), no.~1, 3--40 (in Russian).



\bibitem{[Lytv02]}
Lytvynov E.,
The square of white noise as a Jacobi f\/ield,
{\it Infin. Dimens. Anal. Quantum Probab. Relat. Top.} {\bf 7} (2004), 619--630,
\href{http://arxiv.org/abs/math.PR/0401370}{math.PR/0401370}.


\bibitem{[Meix34]}
Meixner J.,
Orthogonale Polynomsysteme mit einen besonderen Gestalt der erzeugenden Funktion,
{\it J. Lond. Math. Soc.} {\bf 9} (1934), 6--13.

\bibitem{Ovando}
Ovando G.,
Four dimensional symplectic Lie algebras,
{\it Beitr\"age Algebra Geom.} {\bf 47} (2006), 419--434,
\href{http://arxiv.org/abs/math.DG/0407501}{math.DG/0407501}.

\bibitem{Partha-Schmidt}
Parthasarathy K.R., Schmidt K.,
Positive def\/inite kernels continuous tensor products and central limit theorems of probability theory, {\it Springer Lecture Notes in Mathematics}, Vol.~272, Springer-Verlag, Berlin~-- New York, 1972.


\bibitem{9}
Pope C.N.,
Lectures on $W$ algebras and $W$ gravity,
Lectures given at the Trieste Summer School in High-Energy Physics, August 1991.

\bibitem{[Prohor05]}
Prokhorenko D.V., Squares of white
   noise, ${\rm SL}(2,{\mathbb C})$ and Kubo--Martin---Schwinger states, {\it Infin. Dimens. Anal. Quantum Probab. Relat. Top.} {\bf 9} (2006),  491--511.


\bibitem{[Snia99]}
\'Sniady P.,
Quadratic bosonic and free white noises,
{\it Comm. Math. Phys.} {\bf 211} (2000), 615--628,
\mbox{\href{http://arxiv.org/abs/math-ph/0303048}{math-ph/0303048}}.



\bibitem{9+}
Wanglai L., Wilson R.,
Central extensions of some Lie algebras,
{\it Proc. Amer. Math. Soc.} {\bf 126} (1998), 2569--2577.


\bibitem{[Zamol85]}
Zamolodchikov A.B.,
Inf\/inite additional symmetries in two-dimensional conformal quantum f\/ield theory,
{\it Teoret. Mat. Fiz.} {\bf 65} (1985), 347--359.

\end{thebibliography}
\end{document}